\documentclass[journal, final, letterpaper, oneside, twocolumn]{IEEEtran}

\usepackage{color}
\usepackage{amssymb}
\usepackage{amsmath}
\usepackage{graphicx}
\usepackage{multirow}
\usepackage{algorithmic}
\usepackage{epstopdf}
\usepackage{framed}
\usepackage[table]{xcolor}

\usepackage[scriptsize]{subfigure}
\graphicspath{{.}}  

\newenvironment{whitebox}[1][0pt]{%
\MakeFramed{\advance\hsize-\width \advance\hsize-#1 \FrameRestore}}%
{\endMakeFramed}

\newtheorem{theorem}{Theorem}

\newtheorem{claim}[theorem]{Claim}

\renewcommand{\thefigure}{\arabic{figure}}

\newcommand{\cc}{{\bf CoCo} }
\newcommand{\cm}{{\bf CoMa} }
\newcommand{\ncm}{{\bf No-CoMa} }
\newcommand{\lp}{{\bf LiPo} }

\newcommand{\lpm}{{\bf LiPo-} }

\newcommand{\nlpm}{{\bf No-LiPo-} }
\newcommand{\nulp}{{\bf No-Un-LiPo-} }

\newcommand{\datvectrv}{\mathbf{X}}
\newcommand{\outvectrv}{\mathbf{Y}}

\newcommand{\entropy}{\mathit{H}}
\newcommand{\mutualinfo}{\mathit{I}}

\newcommand{\cN}{{\mathcal{N}}}
\newcommand{\cD}{{\mathcal{D}}}
\newcommand{\cT}{{\mathcal{T}}}
\newcommand{\cS}{{\mathcal{S}}}
\newcommand{\cO}{{\mathcal{O}}}

\newcommand{\test}{T}

\newcommand{\const}[1]{{c_{#1}}}

\newcommand{\universe}{n}
\newcommand{\bl}{\universe}

\newcommand{\defective}{d}
\newcommand{\dfct}{\defective}
\newcommand{\dbnd}{D}

\newcommand{\datvect}{\mathbf{x}}
\newcommand{\bx}{\datvect}
\newcommand{\by}{\mathbf{y}}
\newcommand{\bz}{\mathbf{z}}
\newcommand{\bm}{\mathbf{m}}

\newcommand{\defprob}{q}
\newcommand{\defp}{\defprob}

\newcommand{\defpn}{q_0}
\newcommand{\defpp}{q_1}
\newcommand{\dilution}{u}
\newcommand{\dil}{\dilution}

\newcommand{\cost}{\esv}

\newcommand{\measurematx}{M}
\newcommand{\mmat}{\measurematx}
\newcommand{\mmatr}{\mathbf{m}}

\newcommand{\matxprob}{p}

\newcommand{\pmat}{\matxprob}

\newcommand{\prt}{\mathbf{\phi}}
\newcommand{\Prt}{\mathbf{\Phi}}

\newcommand{\Real}{\mathbb{R}}

\newcommand{\esv}{\eta}
\newcommand{\be}{\mathbf{\esv}}

\newcommand{\brd}{\bar{\dfct}}

\newcommand{\threshold}{\tau}
\newcommand{\error}{\epsilon}
\newcommand{\logerror}{\delta}

\IEEEoverridecommandlockouts

\begin{document}

\title{Non-adaptive Group Testing: Explicit Bounds and Novel Algorithms}
%

\author{Chun~Lam~Chan, Sidharth~Jaggi,~\IEEEmembership{Member,~IEEE,} Venkatesh~Saligrama,~\IEEEmembership{Senior~Member,~IEEE,} and\\Samar Agnihotri, ~\IEEEmembership{Member,~IEEE}%
\thanks{The results in this paper were presented in part at the 49th Annual Allerton Conference on Communication, Control, and Computing (Allerton), Oct. 2011~\cite{ChaCJS:11}, and in part at the International Symposium on Information Theory (ISIT) 2012~\cite{ChaJSA:12}.}
\thanks{C. L. Chan and S. Jaggi are with the Chinese University of Hong Kong, Shatin N. T., Hong Kong (e-mail: \{clchan.eric@cuhk.edu.hk, jaggi@ie.cuhk.edu.hk\}).}
\thanks{V. Saligrama is with the Boston University, Boston, MA 02215 USA (e-mail: srv@bu.edu).}
\thanks{S. Agnihotri is with the Indian Institute of Technology Mandi, Mandi, H.P. 175001, India (e-mail: samar@iitmandi.ac.in). He was involved with this work during his stay at the Chinese University of Hong Kong during 2010-12.}
\thanks{Copyright (c) 2014 IEEE. Personal use of this material is permitted.  However, permission to use this material for any other purposes must be obtained from the IEEE by sending a request to pubs-permissions@ieee.org.}
}

\maketitle

\begin{abstract}
We consider some computationally efficient and provably correct algorithms with near-optimal sample-complexity for the problem of noisy non-adaptive group testing. Group testing involves grouping arbitrary subsets of items into pools. Each pool is then tested to identify the defective items, which are usually assumed to be ``sparse''. We consider {\it non-adaptive randomly pooling measurements}, where pools are selected randomly and independently of the test outcomes. We also consider a model where noisy measurements allow for both some false negative and some false positive test outcomes (and also allow for asymmetric noise, and activation noise). 
{We consider three classes of algorithms for the group testing problem (we call them specifically the ``Coupon Collector Algorithm'', the ``Column Matching Algorithms'', and the ``LP Decoding Algorithms'' -- the last two classes of algorithms (versions of some of which had been considered before in the literature) were inspired by corresponding algorithms in the Compressive Sensing literature.
The second and third of these algorithms have several flavours, dealing separately with the noiseless and noisy measurement scenarios. Our contribution is novel analysis to derive explicit sample-complexity bounds -- with all constants expressly computed -- for these algorithms as a function of the desired error probability; the noise parameters; the number of items; and the size of the defective set (or an upper bound on it). We also compare the bounds to information-theoretic lower bounds for sample complexity based on Fano's inequality and show that the upper and lower bounds are equal up to an explicitly computable universal constant factor (independent of problem parameters).}
\end{abstract}


\IEEEpeerreviewmaketitle

\section{Introduction}
\label{sec:intro}
The goal of {\it group testing} is to identify a small unknown subset ${\cD}$ of defective items embedded in a much larger set $\cN$ (usually in the setting where $\dfct=|{\mathcal D}|$ is much smaller than $\bl=|{\cN}|$, {\it i.e.}, $\dfct$ is $o(\bl)$). This problem was first considered by Dorfman \cite{Dorfman1943} in scenarios where multiple items in a group can be simultaneously tested, with a positive or negative output depending on whether or not a ``defective'' item is present in the group being tested. In general, the goal of group testing algorithms is to identify the defective set with as few measurements as possible. As demonstrated in~\cite{Dorfman1943} and later work (see~\cite{CGT} for a comprehensive survey of many group-testing algorithms and bounds), with judicious grouping and testing, far fewer than the trivial upper bound of $\bl$ tests may be required to identify the set of defective items.

We consider the problem of {\it non-adaptive group testing} in this paper (see, for example,~\cite{CGT}). In non-adaptive group testing, the set of items being tested in each test is required to be independent of the outcome of every other test. This restriction is often useful in practice, since this enables parallelization of the testing process. It also allows for an automated testing process using ``off-the-shelf'' components. In contrast, the procedures and hardware required for {\it adaptive} group testing may be significantly more complex.

In this paper we describe computationally efficient algorithms with near-optimal performance for ``noiseless'' and ``noisy'' non-adaptive group testing problems. We now describe different aspects of this work in some detail.

\noindent
{\bf ``Noisy" measurements:}
In addition to the {\it noiseless} group-testing problem ({\it i.e.} wherein each test outcome is ``true''), we consider the {\it noisy} variant of the problem. In this noisy variant the result of each test may differ from the true result (in an independent and identically distributed manner) with a certain pre-specified probability $\defp$. This leads to both {\it false positive test outcomes} and {\it false negative test outcomes}. 

Much of the existing work either considers one-sided noise, namely false positive tests but no false negative tests or vice-versa ({\it e.g.}~\cite{saligrama09}), or an {\it activation noise} model wherein a defective item in a test may behave as a non-defective item with a certain probability (leading to a false negative test) ({\it e.g.}~\cite{saligrama09}),
or a ``worst-case'' noise model~\cite{Macula1997217} wherein the total number of false positive and negative test outcomes are assumed to be bounded from above.\footnote{For instance~\cite{Macula1997217} considers group-testing algorithms that are resilient to {\it all} noise patterns wherein at most a fraction $\defp$ of the results differ from their true values, rather than the probabilistic guarantee we give against {\it most} fraction-$\defp$ errors. This is analogous to the difference between combinatorial coding-theoretic error-correcting codes (for instance Gilbert-Varshamov codes~\cite{GV}) and probabilistic information-theoretic codes (for instance~\cite{Shannon:48}). 
In this work we concern ourselves only with the latter, though it is possible that some of our techniques can also be used to analyzed the former.} Since the measurements are noisy, the problem of estimating the set of defective items is more challenging.\footnote{{\bf We wish to highlight the difference between {\it noise} and {\it errors}. We use the former term to refer to noise in the outcomes of the group-test, regardless of the group-testing algorithm used. The latter term is used to refer to the error due to the estimation process of the group-testing algorithm. Our models allow for noise to occur in measurements, {\it and} also allow for (arbitrarily small) probability of error -- indeed, the latter is a design parameter of our algorithms.}}
{In this work we focus primarily on noise models wherein a positive test outcome is a false positive with the same probability as a negative test outcome being a noisy positive. We do this primarily for ease of analysis (as, for instance, the Binary Symmetric Channel noise model is much more extensively studied in the literature than other less symmetric models), though our techniques also work for asymmetric error probabilities, and for {\it activation} noise. We briefly show in Theorems~\ref{thm:asym} and~\ref{thm:act} that our LP decoding algorithms also have order-optimal sample complexity for these noise models.}

\noindent
{\bf Computationally efficient and near-optimal algorithms:} Most algorithms in the literature focus on optimizing the number of measurements required -- in some cases, this leads to algorithms that may not be computationally efficient to implement (for {\it e.g.}~\cite{saligrama09}).  In this paper we consider algorithms that are not only computationally efficient but are also near-optimal in the number of measurements required. {Variants of some of these algorithms have been considered before in the rich and vast literature on group testing. When we later discuss the specific algorithms considered in this work we also cite related algorithms from the existing literature.}

{We reprise lower bounds on group-testing algorithms based on information-theoretic analysis (similar bounds were known in the literature, for instance~\cite{Mal78, Dya:04} -- we reproduce short proofs here for completeness, since they help us compare upper and lower bounds and show that they match up to a universal constant factor)}. 

For the upper bounds we analyze three different types of algorithms. Some of these algorithms also have structural relationships to those described in the compressive sensing literature (see Section~\ref{subsec:cs}, and our descriptions of the algorithms in Section~\ref{subsec:algorithm}). 

The first algorithm is based on the classical Coupon Collector's Problem~\cite{Chernoff}. The second is based on an iterative ``Column Matching'' algorithm. Hence we call these two algorithms respectively the \cc algorithm (for Coupon Collector),  and the \cm algorithms (for Column Matching algorithms). Neither \cc nor \cm are entirely new -- variants of both have also been previously considered in the group-testing literature (under different names) for both noiseless and noisy scenarios ({see, for instance~\cite{ChenH:08} for algorithms based on the idea of identifying all the non-defectives}, as in the Coupon Collector's algorithm, and~\cite{NGT, MalM:80, MalS:98} for examples of Column Matching algorithms).

Our third class of algorithms are related to linear programming relaxations used in the compressive sensing literature. In compressive sensing the $\ell_0$ norm minimization is relaxed to an $\ell_1$ norm minimization. In the noise-free case this relaxation results in a linear program since the measurements are linear. In contrast, in group testing, the measurements are non-linear and boolean. Nonetheless, our contribution here is to show (via novel ``perturbation analysis'' that)  a variety of relaxations result in a class of LP decoding algorithms that also have information-theoretically order-optimal sample complexity.

In our LP relaxation, in the noise-free case the measurements take the value one if {\it some} defective item is in the pool and zero if no defective item is part of the pool. Furthermore, noise in the group testing scenario is also boolean unlike the additive noise usually considered in the compressive sensing literature. For these reasons we also need to relax our boolean measurement equations. We do so by using a novel combination of inequality and positivity constraints, resulting in a class of Linear Programs.

We first consider the case when the number of defectives $\dfct$ is known exactly (and later relax this constraint). Our LP formulation and analysis is related to error-correction~\cite{McWilliamsS:77} where one uses a ``minimum distance" decoding criteria based on perturbation analysis. The idea is to decode to a vector pair consisting defective items, $\bx$, and the error vector, $\be$, such that the error-vector $\be$ is as ``small" as possible. We call this algorithm the \nlpm algorithm (for Noisy LP). Using novel ``perturbation analysis'', certain structural properties of our LP, convexity arguments, and standard concentration results we show that the solution to our LP decoding algorithm recovers the true defective items with high probability. Based on this analysis, we can directly derive the performance of two other LP-based decoding algorithms. In particular, \lp considers the noiseless measurement scenario. 
Also, for the case when only an upper bound $\dbnd$ on $\dfct$ is known (rather than the explicit value of $\dfct$) we have a decoding algorithm \nulp (for Noisy Universal LiPo). This comprises of a sequence of LPs, one for each positive integer less than $\dbnd$, and we show that with high probability only the LP corresponding to the correct value $\dfct$ outputs a ``meaningful" answer.

Furthermore, we demonstrate that each of these three types of algorithms achieve near-optimal performance in the sense that the sample complexity of our algorithms match information-theoretic lower bounds within a constant factor, where the constant is independent of the number of items $\bl$ and the defective set size $\dfct$ (and at the cost of additional slackness, can even be made independent of the noise parameter $\defp$ and the error probability $\error$). 
Our sample-complexity bounds here are expressly calculated, and all of the constants involved are made precise.

\noindent
{\bf ``Small-error'' probability $\epsilon$}: Existing work for the non-adaptive group testing problem has considered both deterministic and random pooling designs~\cite{CGT}. In this context both deterministic and probabilistic sample complexity bounds for the number of measurements $\test$ that lead to exact identification of the defective items (with high probability ({\it e.g.},~\cite{saligrama09, DyaR:89}), {or with probability $1$ ({\it e.g.},~\cite{DyaRyk82,ErdosFF:82}) have been derived}. These sample complexity bounds in the literature are often asymptotic in nature, and describe the scaling of the number of measurements $\test$ with respect to the number of items $\bl$ and the number of defectives $\dfct$, so as to ensure that the error probability approaches zero. To gain new insights into the constants involved in the sample-complexity bounds we admit a small but fixed error probability, $\error$.\footnote{{As shown in~\cite{DyaRyk82, Dyachkov1989, Furedi:96}}, the number of measurements required to guarantee zero-error reconstruction even in the noiseless non-adaptive group-testing setting scales as $\Omega(\dfct^2\log(\bl)/\log(\dfct))$. Hence our relaxation to the ``small-error'' setting, where it is known in the literature that a number of tests that scales as $\cO(\dfct \log(\bl))$ suffices.} With this perspective we can compare upper and lower bounds for computationally efficient algorithms that hold not only in an order-wise sense, but also where the constants involved in these order-wise bounds can be made explicit, for all values of $\dfct$ and $\bl$.

\noindent
{\bf Explicit Sample Complexity Bounds}:
Our sample complexity bounds are of the form $\test \geq \beta(\defp,\error) \dfct\log(\bl)$. 
The function $\beta(\defp,\error)$ is an explicitly computed function of the noise parameter $\defp$ and admissible error probability $\error$. In the literature, order-optimal upper and lower bounds on the number of tests required are known for the problems we consider (for instance~\cite{saligrama09}, \cite{Sejdinovic2010}). In both the noiseless and noisy variants, the number of measurements required to identify the set of defective items is known to be $\test = \Theta(\dfct\log(\bl))$ -- here $\bl = |\cN|$ is the total number of items and $\dfct = |\cD|$ is the size of the defective subset. In fact, if only $\dbnd$, an upper bound on $\dfct$, is known, then $\test = \Theta(\dbnd\log(\bl))$ measurements are also known to be necessary and sufficient. In most of our algorithms we explicitly demonstrate that we require only a knowledge of $\dbnd$ rather than the exact value of $\dfct$ (in the LP-based algorithms, for ease of exposition we first consider the case when $\dfct$ is known, and then demonstrate in the algorithm \nlpm that this knowledge is not needed). Furthermore, in the noisy variant, we show that the number of tests required is in general a constant factor larger than in the noiseless case (where this constant $\beta$ is independent of both $\bl$ and $\dfct$, but may depend on the noise parameter $\defp$ and the allowable {\it error-probability} $\error$ of the algorithm -- at the cost of additional slackness in the constant even these dependencies can be removed).\footnote{In fact, while our analysis can be done for all values of $\dfct$ and $\bl$, in situations where preserving such generality makes the bounds excessively unwieldy we sometimes use the fact that $\dbnd = o(\bl)$, or that $\dfct = \omega(1)$.}

\subsection{Our contributions}
We summarize our contributions in this work as follows:
{
\begin{itemize}
\item Our contribution for the \cc algorithm is that the analysis draws a simple but interesting connection to  the classical coupon-collector problem. As far as we are aware this connection is novel, and hence interesting in its own right.

\item Our contribution for \cm algorithms is an explicit analysis of the sample complexity, in comparison to the previous literature for such algorithms, which typically focused on order-optimal analysis. {Some elegant work in the Russian literature also has information-theoretically optimal analysis for high-complexity decoding for related algorithms~\cite{DyaR:89, Dya:04}, and good bounds~\cite{MalM:80,MalS:98} for lower-complexity variants (called the ``Separate Testing of Inputs'' algorithm) of these \cm algorithms. However the types of statistical decoding rules used therein are structurally different than the ones we analyze, which can be implemented as dot-products of vectors of real numbers, and hence are particularly computationally efficient in modern computer architectures.}

\item Our contribution for our suite of novel\footnote{In the noiseless setting prior work by~\cite{MalS:09, Mal:10} had considered a similar LP approach (though no detailed analysis was presented). For the scenario with noisy measurements, in parallel with the conference version of our work presented at~\cite{ChaJSA:12}, a similar algorithm was considered in~\cite{MalM:12} -- however, the proof sketch presented therein requires {\it disjunct} measurement matrices, which are known not to be order-optimal in sample-complexity in the ``small-error'' case~\cite{Mal78,Dya:04}.} LP algorithms is a novel perturbation analysis demonstrating that they are indeed order-optimal in sample-complexity. We also demonstrate the versatility of these algorithms and our techniques for analyzing them, by demonstrating that:\\
\item There are multiple LP relaxations for the underlying non-linear group-testing problem, each demonstrating interesting properties. For instance, we demonstrate via the \nlpm decoding algorithm that {\it just} the negative test outcomes alone carry enough information to be able to reconstruct (with high probability) the input $\bx$ with order-optimal sample-complexity, even in the presence of measurement noise. 
In the scenario with no noise, the \lpm decoding algorithm takes a particularly simple form, where it reduces to solving just a feasibility problem. These different relaxations might be of interest in different scenarios. In addition, our analytical techniques directly generalize to other noise models, such as asymmetric noise, and activation noise.

\item The upper bounds we present differ from information-theoretic lower bounds by a constant factor that is independent of problem parameters.

\item Also, many of the outer bounds we present are in the main {\it universal}. That is, most of the classes of algorithms we consider (except for those based on Linear Programming -- universal versions of these algorithms is a subject of ongoing research) do not need to know the exact values of $\dfct$ -- as long as an outer bound $\dbnd$ is known, this suffices.\footnote{{Prior work, for instance by~\cite{GilIS:08}, had also considered group testing that was similarly universal in $\dbnd$.}}

\item We consciously draw connections between the compressive sensing and the group-testing literature, to show that despite the differences in models (linear versus non-linear, real measurements versus real measurements, {\it etc.}) algorithms that work for one setting may well work in another. Such a connection has indeed been noted before -- for instance,~\cite{CorM:06} used group-testing ideas to construct compressive sensing algorithms. Also,~\cite{MalS:09, Mal:10, MalM:12} were motivated by CS-type LP-decoding algorithms to consider corresponding decoding algorithms for the group-testing problem.

\end{itemize}
}

This paper is organized as follows. In Section \ref{sec:Background}, we introduce the model and corresponding notation (frequently used notation is summarized in Table~\ref{table:notation}). In separate subsections of this section we give background on some compressive sensing algorithms, describe the algorithms analyzed in this work, and the relationships between these algorithms for these two disparate problems. We also reprise known information-theoretic lower bounds on the number of tests required by any group-testing algorithm (with short proof sketches in the Appendix). In Section~\ref{sec:Result}, we describe the main results of this work. Section~\ref{sec:uprbnd} contains the analysis of the group-testing algorithms considered. 

\section{Background}
\label{sec:Background}

\subsection{Model and Notation}

\begin{figure}[!t]
	\centering
		\includegraphics[width=90mm]{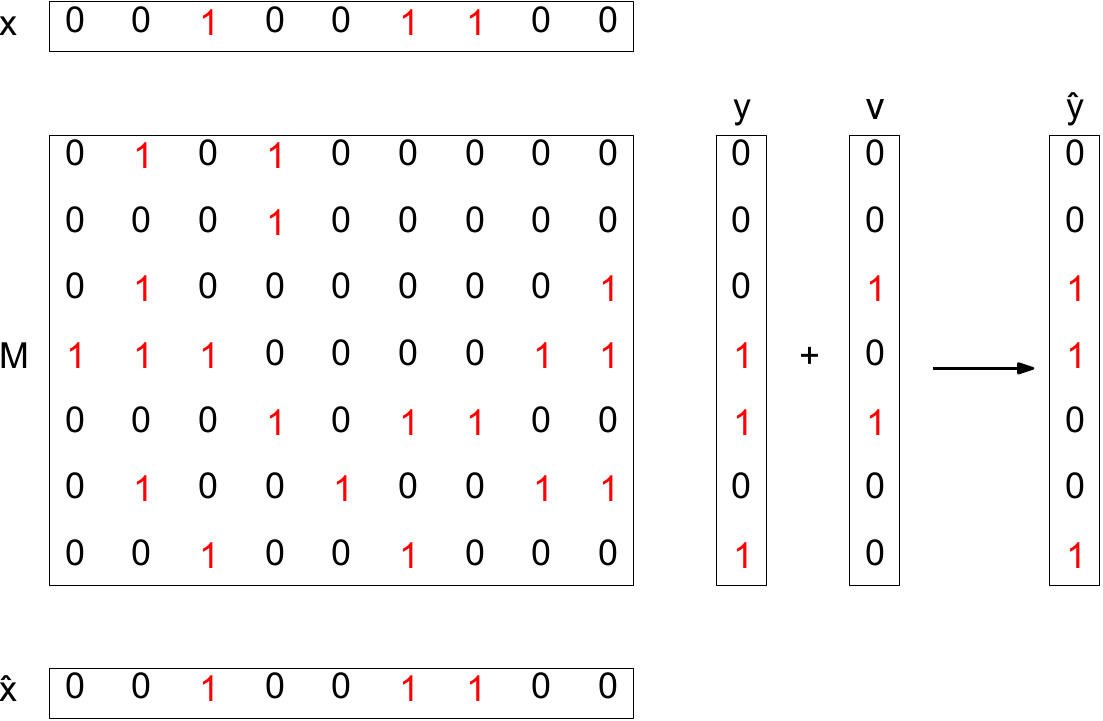}
		\caption {An example demonstrating a typical non-adaptive group-testing setup. The $\test \times \bl$ binary group-testing matrix represents the items being tested in each test, the length-$\bl$ binary input vector ${\bx}$ is a weight $\dfct$ vector encoding the locations of the $\dfct$ defective items in ${\cD}$, the length-$\test$ binary vector ${\mathbf y}$ denotes the outcomes of the group tests in the absence of noise, the length-$\test$ binary noisy result vector $\hat{\by}$ denotes the actually observed noisy outcomes of the group tests, as the result of the noiseless result vector being perturbed by the length-$\test$ binary noise vector ${\mathbf \nu}$. The length-$\bl$ binary estimate vector $\hat{\bx}$ represents the estimated locations of the defective items.}
		\label{fig:gt}
\end{figure}

A set ${\cN}$ contains $\bl$ items, of which an unknown subset ${\cD}$ are said to be ``defective". 
{{The size of $\cD$ is denoted by $\dfct$. We assume that $\dbnd$, an upper bound on the true value of $\dfct$, is known {\it a priori}.} }
\footnote{It is common (see for example \cite{springerlink:10.1007/BF01609876}) to assume that the number $\dfct$ of defective items in ${\cD}$ is known, or at least a good upper bound $\dbnd$ on $\dfct$, is known {\it a priori}. If not, other work (for example \cite{SOBEL01041975}) considers non-adaptive algorithms with low query complexity that help estimate $\dfct$. However, in this work we explicitly consider algorithms that do not require such foreknowledge of $\dfct$ -- rather, our algorithms have ``good" performance with ${\cal O}(\dbnd\log(\bl))$ measurements.} The goal of group-testing is to correctly identify the set of defective items via a minimal number of ``group tests", as defined below (see Figure~\ref{fig:gt} for a graphical representation).

Each row of a $\test \times \bl$ binary {\it group-testing matrix} $\mmat$ corresponds to a distinct test, and each column corresponds to a distinct item. Hence the items that comprise the group being tested in the $i$th test are exactly those corresponding to columns containing a $1$ in the $i$th location. The method of generating such a matrix $\mmat$ is part of the design of the group test -- this and the other part, that of estimating the set ${\cD}$, is described in Section~\ref{subsec:algorithm}.

The length-$\bl$ binary {\it input} vector $\bx$ represents the set ${\cN}$, and contains $1$s exactly in the locations corresponding to the items of ${\cD}$. The locations with ones are said to be {\it defective} -- the other locations are said to be {\it non-defective}. We use these terms interchangeably.

The outcomes of the {\it noiseless} tests correspond to the length-$\test$ binary {\it noiseless result} vector ${\by}$, with a $1$ in the $i$th location if and only if the $i$th test contains at least one defective item.

The observed vector of test outcomes in the {\it noisy} scenario is denoted by the length-$\test$ binary {\it noisy result} vector $\hat{\by}$ -- the probability that each entry $y_i$ of ${\by}$ differs from the corresponding entry $\hat{y}_i$ in $\hat{\by}$ is $\defp$, where $\defp$ is the {\it noise parameter}. The locations where the noiseless and the noisy result vectors differ is denoted by the length-$\test$ binary {\it noise vector} ${\mathbf \nu}$, with $1$s in the locations where they differ. 
{Tests (if any) in which the noise perturbs a negative to a positive outcome are called {\it false positives}, and tests (if any) in which the noise perturbs a positive to a negative outcome are called {\it false negatives}.

Two alternate noise models are also briefly considered in Theorems~\ref{thm:asym} and~\ref{thm:act}. Firstly, the {\it asymmetric bit-flip} model generalized the BSC($\defp$) model described in the previous paragraph, by allowing the probabilities of 
false positive test outcomes and false negative test outcomes (denoted respectively by $\defpn$ and $\defpp$) to be different. Secondly, in the {\it activation noise} model individual defective items in tests have a {\it non-activation} probability $\dil$. That is, defective items in a test have an i.i.d. probability of behaving like non-defectives in that test. This leads to false negative test outcomes. In addition, in the activation noise model we {\it also} allow false positive test outcomes to occur in an i.i.d. manner (with probability $\defpn$). This is a generalization of the ``usual'' activation noise model (for instance~\cite{saligrama09}). 
}

The estimate of the locations of the defective items is encoded in the length-$\bl$ binary {\it estimate vector} $\hat{\bx}$, with $1$s in the locations where the group-testing algorithms described in Section~\ref{subsec:algorithm} estimate the defective items to be.
{Items (if any) in which non-defective items are (incorrectly) decoded as defective items are called {\it false defective estimates}, and items (if any) in which defective items are (incorrectly) decoded as non-defective items are called a {\it false non-defective estimate}.}

The {\it probability of error} of any group-testing algorithm is defined as the probability (over the input vector ${\bx}$  and noise vector ${\mathbf{\nu}}$) that the estimated vector differs from the input vector at all.
\footnote{However, as is common in information-theoretic proofs, we instead calculate the probability of error for any fixed $\bx$ averaged over the randomness in the noise vector ${\mathbf{\nu}}$ and over the choice of measurement matrices $\mmat$ (which are chosen from a random ensemble). Standard averaging arguments then allow us to transition to the definition of probability of error stated above. Since this step is common to each of our proofs, we do not explicitly state this henceforth.}

\begin{table*}[!t]
\centering
\begin{tabular}{c | l }
\hline
${\cN}$ & The set of all items being tested.\\
$\bl$ & The total number of items, $n = |{\mathcal N}|$.\\
${\cD}$ & The unknown set of all defective items.\\
$\dfct$ & The total number of defective items, $d = |{\mathcal D}|$.\\
$\dbnd$ & The upper-bound on $d$.\\
$\test$ & The number of measurements required to identify the set of defective items.\\
$\mmat$ & The $\test \times \bl$ binary group-testing matrix.\\
$\pmat$ & Probability with which each element in $\mmat$ is chosen as $1$.\\
${\bx}$ & The length-$\bl$ binary input vector $(x_1,\ldots,x_\bl)$ that is a weight $\dfct$ vector encoding the locations of the $\dfct$ defective items in ${\cD}$.\\
${\by}$ & The length-$\test$ binary vector $(y_1,\ldots,y_\bl)$ denoting the outcomes of the group tests in the absence of noise.\\
$\defp$ & The pre-specified probability that the result of a test differs from the true result.\\
$\defpp$ & In the asymmetric noise model, the probability of a false negative test outcome.\\
$\defpn$ & In the asymmetric and activation noise models, the probability of a false positive test outcome.\\
$\dil$ & In the activation noise model, the probability of the activation noise of a defective item in a test\\
${\mathbf \nu}$ & The length-$\test$ binary noise vector $(\nu_1,\ldots,\nu_\bl)$.\\
$\hat{\by}$ & The length-$\test$ binary noisy result vector that denotes the actually observed noisy outcomes of the group tests, equaling $\hat{\by} = \by + \nu$.\\
$\hat{\bx}$ & The length-$\bl$ binary estimate vector that represents the estimated locations of the defective items.\\
$\error = \bl^{-\logerror}$ & The pre-specified small but fixed error probability of the algorithm.\\
$\threshold$ & A decoding threshold for the NCOMP algorithm.\\
$\Gamma = \frac{\ln \dbnd}{\ln \bl}$ & A measure of how much smaller $\cD$ is than $\cN$. Takes values between $0$ and $1$\\
$\gamma = \frac{\Gamma + \delta}{1 + \delta}$ & An internal parameter defining the performance of many of our algorithms.\\
$g$ & The number of items selected to form a group in CBP.\\
$\cost_i$ & In our LP-based algorithms, the ``slack variable'' to account for errors in the $i^\textrm{th}$ test outcome, $i \in \{1, \ldots, \test\}$.\\
$\prt'$ & In our LP-based algorithms, a perturbation vector.\\
$\Prt'$ & In our LP-based algorithms, the set of perturbation vectors.\\
$\Delta'_{0,i}$ & In our LP-based algorithms, the expected change in the objective value of the LP when $\bx$ is perturbed to $\bx+\prt'$.\\
\hline
\end{tabular}
\vspace{0.1in}
\caption{Notation used frequently in the paper. The asymmetric noise and activation noise models are only briefly considered in Theorems~\ref{thm:asym} and~\ref{thm:act}. Of the last five items, last four are internal variables specifically in our LP-based algorithms, and the other ($g$) is an internal variable specifically in our Coupon-Collector model. All other notation is ``universal'' (applies to the entirety of the paper).}
\label{table:notation}
\end{table*}

\subsection{Compressive Sensing}\label{subsec:cs}
Compressive sensing has received significant attention over the last decade. We describe the version most related to the topic of this paper~\cite{Candes2008, BaranuikDavDeVWak2008}. This version considers the following problem. Let $\bx$ be an {\it exactly $\dfct$-sparse} vector in $\Real^\bl$, {\it i.e.}, a vector with at most $\dfct$ non-zero components (in general in the situations of interest $\dfct = o(\bl)$).\footnote{As opposed to an {\it approximately $\dfct$-sparse} vector, {\it i.e.}, a vector such that ``most" of its energy is confined to $\dfct$ indices of $\bx$. One way of characterizing such vectors is to say that $||\bx-\bx_\dfct||_1 \leq \const{1} ||\bx ||_1$ for some suitably small $0 < \const{1} < 1$. Here $\bx_\dfct$ is defined as the vector matching $\bx$ on the $\dfct$ components that are largest in absolute value, and zero elsewhere. The results of~\cite{Candes2008, BaranuikDavDeVWak2008} also apply in this more general setting. However, in this work we are primarily concerned with the problem of group testing rather than that of compressive sensing, and present the work of compressive sensing merely by way of analogy. Hence we focus on the simplest scenarios in which we can draw appropriate correspondences between the two problems.}

Let $\bz$ corresponds to a {\it noise vector} added to the measurement $\mmat \bx$.
One is given a set of ``compressed noisy measurements" of $\bx$ as
\begin{eqnarray}
\by & = & \mmat \bx + \bz \label{eq:cs1}\\
||\bz||_2 & \leq & \const{2}\label{eq:cs2}\\
||\bx||_0 & \leq & \dfct \label{eq:cs3}
\end{eqnarray}
\noindent
Here the constraint (\ref{eq:cs2}) corresponds to a guarantee that the noise is not ``too large", and the (non-linear) constraint (\ref{eq:cs3}) corresponds to the prior knowledge that $\bx$ is $\dfct$-sparse.
The $\test \times \bl$ matrix $\mmat$ is designed by choosing each entry i.i.d. from a suitable probability distribution (for instance, the set of zero-mean, $1/\bl$ variance Gaussian random variables). The decoder must use the resulting {\it noisy measurement vector} $\by \in \Real^\test$ and the matrix $\mmat$ to computationally efficiently estimate the underlying vector $\bx$. The challenge is to do so with as few measurements as possible, {\it i.e.}, with the number of rows $\test$ of $\mmat$ being as small as possible.

We now outline two well-studied decoding algorithms for this problem, each of which motivated one class of algorithms we analyze in this work.

\subsubsection{Orthogonal Matching Pursuit}
We note that it is enough for the decoder to computationally efficiently estimate the {\it support $\cD$ of $\bx$}, the set of indices on which $\bx$ is non-zero, correctly.
This is because the decoder can then estimate $\bx$ as $(\mmat_\cD^t\mmat_\cD)^{-1}\mmat_\cD^t\by$, which is the minimum mean-square error estimate of $\bx$. (Here $\mmat_\cD$ equals the $\test \times \dfct$ sub-matrix of $\mmat$ whose columns numbers correspond to the indices in $\cD$, and $\test$ is a design parameter chosen to optimize performance.)

One popular method of efficient estimation of $\cD$ is that of {\it Orthogonal matching pursuit} (OMP) \cite{Tropp07signalrecovery}\footnote{{In fact, there are {\it many} techniques that have a similar ``column matching'' flavour -- see, for instance~\cite{DonTDS:12,DaiM:09,NeeT:10}. We focus on OMP just for its simplicity.}}. The intuition is that if the columns of the matrix $\mmat$ are ``almost orthogonal" (every pair of columns have ``relatively small" dot-product) then decoding can proceed in a greedy manner. In particular, the OMP algorithm computes the dot-product between $\by$ and each column $\bm_i$ of $\mmat$, and declares $\cD$ to be the set of $\dfct$ indices for which this dot-product has largest absolute value.

One can show~\cite{Tropp07signalrecovery} that there exists a universal constant $\const{3}$ such that if $\bz = 0$ then with probability at least $1-\dfct^{-\const{3}}$ (over the choice of $\mmat$, which is assumed to be independent of the vector $\bx$) this procedure correctly reconstructs $\bx$ with $T \leq \const{3}\dfct\log(\bl)$ measurements. Similar results can also be shown with $\bz \neq 0$, though the form of the result is more intricate.

\subsubsection{Basis Pursuit}
An alternate decoding procedure proceeds by relaxing the compressive sensing problem (in particular the non-linear constraint (\ref{eq:cs3})) into the convex optimization problem called {\it Basis Pursuit} (BP).
\begin{eqnarray}
\bx = \arg\min ||\bx||_1 \label{eq:bp}\\
\mbox{subject to }
||\by - \mmat \bx||_2 \leq \const{2}\label{eq:bp1}
\end{eqnarray}

It can be shown (for instance~\cite{Candes2008, BaranuikDavDeVWak2008}) that there exist constants $\const{4}, \const{5}$ and $\const{6}$ such that with $\test = \const{4}\dfct\log(\bl)$, with probability at least $1-2^{\const{5}\bl}$, the solution $\bx^*$ to BP satisfies $||\bx^* - \bx||_2 \leq \const{6}||\bz||_2$.

\subsection{Group-Testing Algorithms in this work}
\label{subsec:algorithm}

We now describe three classes of algorithms (in the latter two cases, in both the noiseless and noisy settings). The algorithms are specified by the choices of encoding matrices and decoding algorithms. Their performance is stated in Section~\ref{sec:Result} and the corresponding proofs of the algorithms are presented in Section~\ref{sec:uprbnd}.

\subsubsection{Coupon collector algorithm (\cc)}
\hspace*{\fill} \\
We first consider an algorithm that consider rows of the measurement matrix $\mmat$, and try to correlate these with the observations $\by$. 
\begin{figure}[!t]
	\centering
		\includegraphics[width=1.5in]{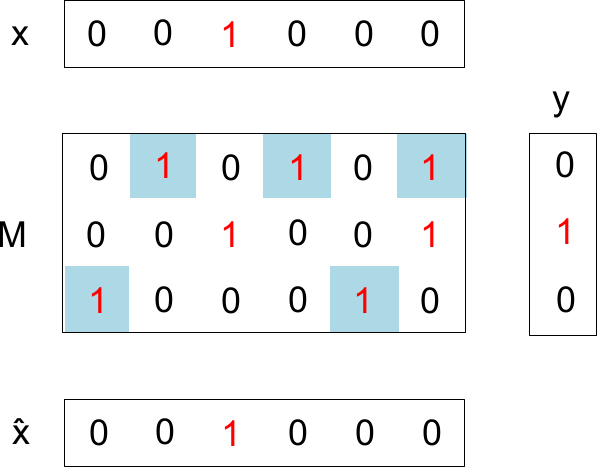}
		\caption {An example demonstrating the CBP algorithm. Based on only on the outcome of the negative tests (those with output zero), the decoder estimates the set of non-defective items, and ``guesses" that the remaining items are defective.}
		\label{fig:cbp}
\end{figure}

The $\test \times \bl$ group-testing matrix $\mmat$ is defined as follows. A {\it group sampling parameter} $g$ is chosen (the exact values of $\test$ and $g$ are code-design parameters to be specified later). Then, the $i$th row of $\mmat$ is specified by sampling with replacement\footnote{Sampling without replacement is a more natural way to build tests, but the analysis is trickier. However, the performance of such a group-testing scheme can be shown to be no worse than the one analyzed here~\cite{Rice:06}. Also see Footnote~\ref{fn:wrt}.} from the set $[1,\ldots,\bl]$ exactly $g$ times, and setting the $(i,j)$ location to be one if $j$ is sampled at least once during this process, and zero otherwise.\footnote{\label{fn:wrt}Note that this process of sampling each item in each test with replacement results in a slightly different distribution than if the group-size of each test was fixed {\it a priori} and hence the sampling was ``without replacement" in each test. (For instance, in the process we define, each test may, with some probability, test fewer than $g$ items.) The primary advantage of analyzing the ``with replacement" sampling is that in the resulting group-testing matrix every entry is then chosen {\it i.i.d.}.}

The decoding algorithm proceeds by using {\it only} the tests which have a negative outcome, to identify all the non-defective items, and declaring all other items to be defective. If $\mmat$ is chosen to have enough rows (tests), each non-defective item should, with significant probability, appear in at least one negative test, and hence will be appropriately accounted for. Errors (false defective estimates) occur when at least one non-defective item is not tested, or only occurs in positive tests ({\it i.e.,} every test it occurs in has at least one defective item). The analysis of this type of algorithm comprises of estimating the trade-off between the number of tests and the probability of error.

More formally, for all tests $i$ whose measurement outcome $y_i$ is a zero, let ${\mmatr_i}$ denote the corresponding $i$th row of $\mmat$. The decoder outputs $\hat{\bx}$ as the length-$\bl$ binary vector which has $0$s in exactly those locations where there is a $1$ in at least one such ${\mmatr_i}$. 


\subsubsection{``Column Matching" Algorithms}
We next consider algorithms that try to correlate columns of the measurement matrix $\mmat$ with the vector of observations $\hat{\by}$. We consider two scenarios -- the first when the observations are noiseless, and the second when they are noisy. In both cases we draw the analogy with a corresponding compressive sensing algorithm.

\noindent {\bf Column Matching algorithm (\cm)}:

\begin{figure}[!t]
	\centering
		\includegraphics[width=3.5in]{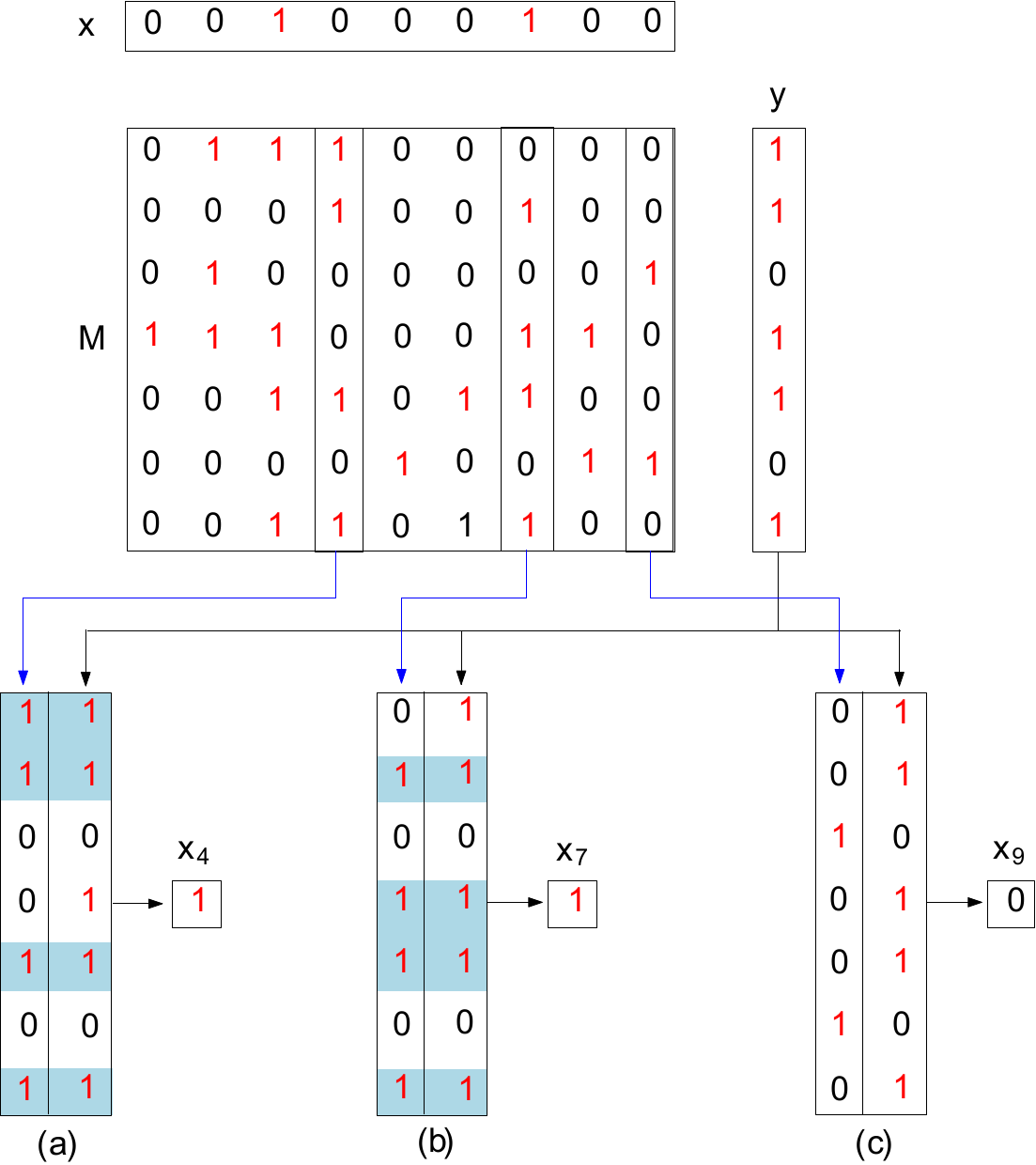}
		\caption {An example demonstrating the \cm algorithm. The algorithm matches columns of $\mmat$ to the result vector. As in (b) in the figure, since the result vector ``contains" the $7$th column, then the decoder declares that item to be defective. Conversely, as in (c), since there is no such containment of the last column, then the decoder declares that item to be non-defective. However, sometimes, as in (a), an item that is truly non-defective, is ``hidden" by some other columns corresponding to defective items, leading to a false defective estimate.}
		\label{fig:comp}
\end{figure}

The $\test \times \bl$ group-testing matrix $\mmat$ is defined as follows. A {\it group selection parameter} $\pmat$ is chosen (the exact values of $\test$ and $\pmat$ are code-design parameters to be specified later). Then, i.i.d. for each $(i,j)$, $m_{i,j}$ (the $(i,j)$th element of $\mmat$) is set to be one with probability $\pmat$, and zero otherwise.

The decoding algorithm works ``column-wise" on $\mmat$ -- it attempts to match the columns of $\mmat$ with the result vector ${\by}$. That is, if a particular column $j$ of $\mmat$ has the property that all locations $i$ where it has ones {\it also} corresponds to ones in $y_i$ in the result vector, then the $j$th item ($x_j$) is declared to be defective. All other items are declared to be non-defective.
\footnote{\label{fn:COMP} Note the similarity between this algorithm and OMP. {Just as OMP tries to select columns of the measurement matrix that have ``small angle'' (large dot-product) with the vector of observations, similarly this algorithm tries to detect columns of $\mmat$ with maximal ``overlap'' with the result vector $\by$. Indeed, even though the measurement process in the group-testing problem is fundamentally non-linear, in current computer architectures where linear algebraic operations often have low computational complexities, a computationally efficient means of implementing such a test is to check whether the dot-product between $\by$ and each column of $\mmat$ equals the number of ones in that column.  
Hence the name \cm.}}

Note that this decoding algorithm never outputs false non-defective estimates, only false defective estimates. A false defective estimate occurs when {\it all} locations with ones in the $j$th column of $\mmat$ (corresponding to a non-defective item $j$) are ``hidden" by the ones of other columns corresponding to defectives items. That is, let column $j$ and some other columns $j_1,\ldots,j_k$ of matrix $\mmat$ be such that for each $i$ such that $m_{i,j} = 1$, there exists an index $j'$ in $\{j_1,\ldots,j_k\}$ for which $m_{i,j'}$ also equals $1$. Then if each of the $\{j_1,\ldots,j_k\}$th items are defective, then the $j$th item will also always be declared as defective by the \cm decoder, regardless of whether or not it actually is. The probability of this event happening becomes smaller as the number of tests $\test$ become larger.

The rough correspondence between this algorithm and Orthogonal Matching Pursuit (\cite{Tropp07signalrecovery}) arises from the fact that, as in Orthogonal Matching Pursuit, the decoder attempts to match the columns of the group-testing matrix with the result vector.

\noindent {\bf Noisy Column Matching algorithm (\ncm)}

\begin{figure}[!t]
	\centering
		\includegraphics[width=3.5in]{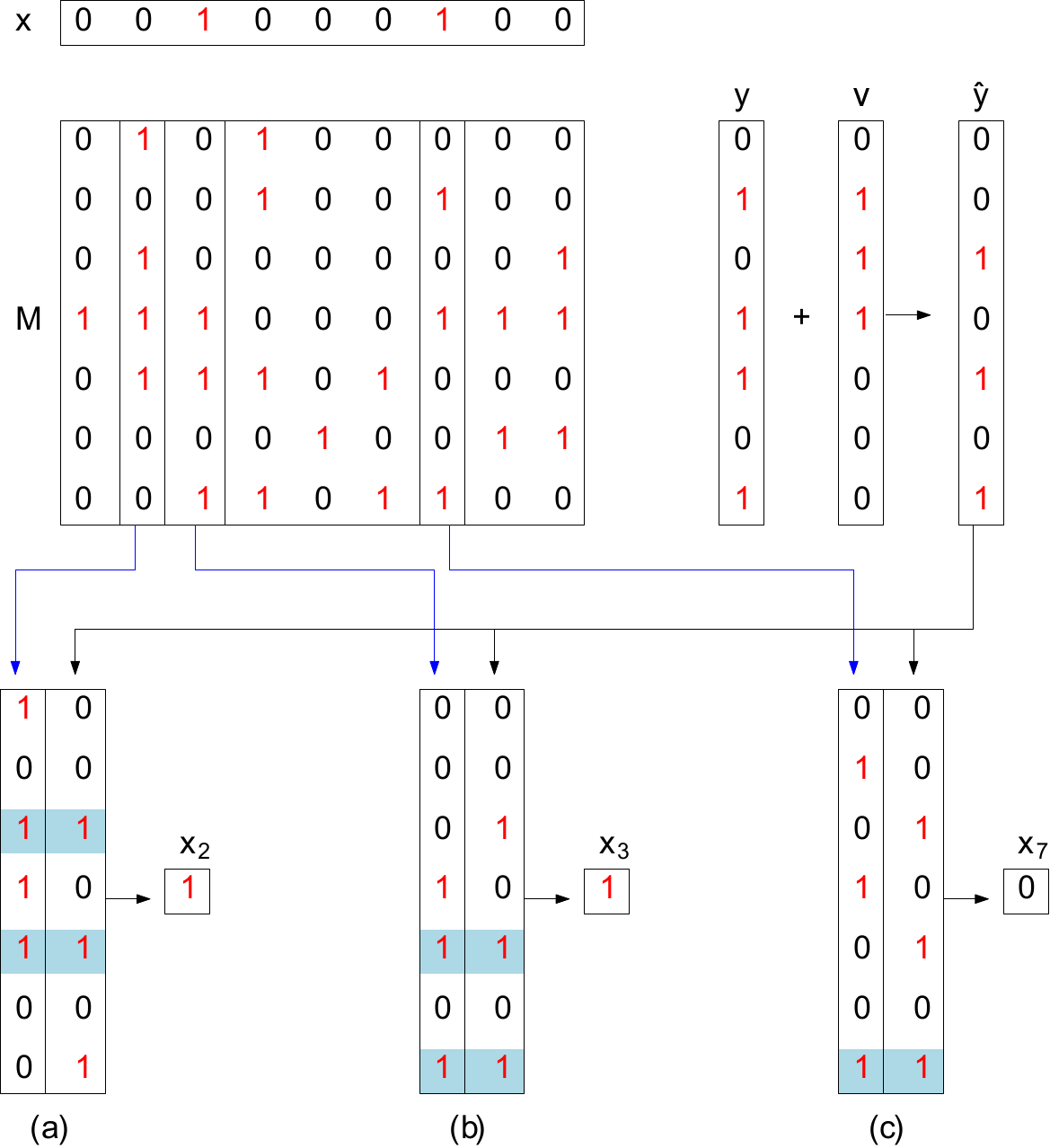}
		\caption {An example demonstrating the \ncm algorithm. The algorithm matches columns of $\mmat$ to the result vector {\it up to a certain number of mismatches} governed by a threshold. In this example, the threshold is set so that the number mismatches be less than the number of matches. For instance, in (b) above, the $1$s in the third column of the matrix match the $1$s in the result vector in two locations (the $5$th and $7$th rows), but do not match only in one location in the $4$th row (locations wherein there are $0$s in the matrix columns but $1$s in the result vector do not count as mismatches). Hence the decoder declares that item to be defective, which is the correct decision. \protect \\ However, consider the false non-defective estimate generated for the item in (c). This corresponds to the $7$th item. The noise in the $2$nd, $3$rd and $4$th rows of ${\mathbf \nu}$ means that there is only one match (in the $7$th row) and two mismatches ($2$nd and $4$th rows) -- hence the decoder declares that item to be non-defective. \protect \\ Also, sometimes, as in (a), an item that is truly non-defective estimate, has a sufficient number of measurement errors that the number of mismatches is reduced to be below the threshold, leading to a false defective estimate. }
		\label{fig:ncomp}
\end{figure}

In the \ncm algorithm, we relax the sharp-threshold requirement in the original \cm algorithm that the set of locations of ones in any column of $\mmat$ corresponding to a defective item be {\it entirely} contained in the set of locations of ones in the result vector. Instead, we allow for a certain number of ``mismatches" -- this number of mismatches depends on both the number of ones in each column, and also the noise parameter $\defp$.

Let $\pmat$ and ${\threshold}$ be design parameters to be specified later. To generate the $\mmat$ for the \ncm algorithm, each element of $\mmat$ is selected i.i.d. with probability $\pmat$ to be $1$.

The decoder proceeds as follows, For each column $j$, we define the {\it indicator set} ${\cT}_j$ as the set of indices $i$ in that column where $m_{i,j} = 1$. We also define the {\it matching set} ${\cS}_j$ as the set of indices $j$ where both $\hat{y}_i = 1$ (corresponding to the noisy result vector) and $m_{i,j}=1$.

Then the decoder uses the following ``relaxed" thresholding rule. If $|{\cS}_j| \geq |{\cT}_j|(1-\defp(1+\threshold))$, then the decoder declares the $j$th item to be defective, else it declares it to be non-defective.\footnote{As in Footnote~\ref{fn:COMP}, this ``relaxed" thresholding rule can be viewed as analogous to corresponding relaxations in OMP-type algorithms when the observations are noisy.}

\subsubsection{LP-decoding Algorithms}
We now consider LP-decoding algorithms in both the noiseless and noisy settings. The algorithms are specified by the choices of encoding matrices and decoding algorithms.
The $\test \times \bl$ group-testing matrix $\mmat$ is defined by randomly selecting each entry in it in an i.i.d. manner to equal $1$ with probability $\pmat = 1/\dbnd$, and $0$ otherwise.


\begin{figure}
\begin{whitebox}
\begin{eqnarray}
(\hat{\bx},\hat{\mathbf{\cost}})=\arg\min_{\bx,\be}\sum_{i} \esv_i \label{eq:ncbp0}\\
\mbox{such that \mbox{\hspace{2in}}}\nonumber\\
- \esv_i + \sum_{j:m_{ij} = 1} x_j = 0, \mbox{ if $\hat{y_i} = 0$,} \label{eq:ncbp1}\\
\sum_{\forall j} x_j =\dfct,  \label{eq:ncbp3}\\
0 \leq x_j \leq 1, \label{eq:ncbp4}\\
0 \leq \esv_i \leq \dfct \label{eq:ncbp5}
\end{eqnarray}
\end{whitebox}
\label{fig:NCBP-LP}
\vspace{-0.2in}
\caption{{\bf \nlpm:} Constraint (\ref{eq:ncbp4}) relaxes the constraint that each $x_j \in \{0,1\}$,  and constraint (\ref{eq:ncbp3}) indicates that there are exactly $\brd$ defective items in the $\brd$th iteration of the LP.
The variables $\esv_i$ are ``slack variables" in the equations (\ref{eq:ncbp1}). For instance, if test $i$ is truly negative, then all the variables in an equation of the form (\ref{eq:ncbp1}) are zero. However, if the test is a false negative, then the variable $\esv_i$ is then set to equal the number of defective items in test $i$. Note that $\esv_i$ is bounded above by $\dfct$ in the case of (false) negatives. 
Note that this linear program only uses negative test outcomes.
In principle, one could also have defined constraints corresponding to positive test outcomes. However, we are unable to analyze the performance of the resulting LP (though simulations indicate that such an LP does index perform well).
}
\end{figure}

\noindent {\bf Noisy Linear Programming decoding (\nlpm)}:

A linear relaxation of the group testing problem with noisy measurements leads ``naturally'' to \nlpm (\ref{eq:ncbp0})-(\ref{eq:ncbp5}).
In particular, each $x_i$ is relaxed to satisfy $0 \leq x_i \leq 1$, and the non-linear measurements are linearized in (\ref{eq:ncbp1}).
Also, we define ``slack" variables $\esv_i$ for all $i \in \{1,\ldots,\test\}$ to account for errors in the test outcome. For a particular test $i$ this $\esv_i$ is defined to be zero if a particular test result is correct, and positive (and at least $1$) if the test result is incorrect. Of course, the decoder does not know {\it a priori} which scenario a particular test outcome falls under, and hence has to also decode $\be$. Nonetheless, as is common in the field of error-correction~\cite{McWilliamsS:77}, often using a ``minimum distance" decoding criteria (decoding to a vector pair $(\bx,\be)$ such that the error-vector $\be$ is as ``small" as possible) leads to good decoding performance. Our LP decoder attempts to do so.
To be precise, the \nlpm decoder  outputs the $(\hat{\bx}, \hat{\cost})$ that minimize the LP given in (\ref{eq:ncbp0})-(\ref{eq:ncbp5}). Note that we assume that the value of $\dfct$ is known precisely in this algorithm 
.

\noindent {\bf Linear Programming decoding (\lp)}:

\begin{figure}
\begin{whitebox}
\begin{eqnarray}
{\hat{\bx}} = \mbox{ feasible point in } \nonumber\label{eq:cbp0}\\
\sum_{j:m_{ij} = 1} x_j = 0, \mbox{ if $y_i = 0$,} \label{eq:cbp1}\\
\sum_{\forall j} x_j = \dfct   \label{eq:cbp3}\\
0 \leq x_j \leq 1 \label{eq:cbp4}
\end{eqnarray}
\end{whitebox}
\label{fig:CBP-LP}
\caption{{\bf \lp:} 
This LP simply attempts to find {\it any} feasible solution for any value of $\brd \in \{0,\ldots,\dbnd\}$
}
\end{figure}

\lp, which analyzes the scenario with noiseless measurements,  is a special case of \nlpm with each $\cost_i$ variable set to zero. Hence it reduces to the problem of finding {\it any} feasible point in the constraint set (\ref{eq:cbp1}-\ref{eq:cbp4})

\subsection{Lower bounds on the number of tests required}

We first reprise statements of information-theoretic lower bounds (versions of which were considered by~\cite{Mal78,Dya:04}) on the number of tests required by {\it any} group-testing algorithm. 
For the sake of completeness, so we can benchmark our analysis of the algorithms we present later, we state these lower bounds here and prove them in the Appendix. 

\begin{theorem}
\label{thm1} \label{thm:1} \label{thm:ITLowBnd}
[\cite{Mal78,Dya:04}] Any group-testing algorithm with noiseless measurements that has a probability of error of at most $\error$ requires at least 
$(1-\bl^{-\logerror})\dbnd \log \left (\frac{\bl}{ \dbnd} \right ) = (1-\bl^{-\logerror})(1-\Gamma)\dbnd \log(\bl)$
\noindent tests.
\end{theorem}

In fact, the corresponding lower bounds can be extended to the scenario with noisy measurements as well.

\begin{theorem}
\label{thm2} \label{thm:2} \label{thm:nlb} \label{thm:ITNoisyLowBnd}
[\cite{Mal78,Dya:04}] Any group-testing algorithm that has measurements that are noisy i.i.d. with probability $\defp$ and that has a probability of error of at most $\error$ requires at least $\frac{(1-\bl^{-\logerror})\dbnd\log\left (\frac{\bl}{ \dbnd} \right )}{1-H(\defp)} = \frac{(1-\bl^{-\logerror})(1-\Gamma)\dbnd\log\left (\bl \right )}{1-H(\defp)}$ tests.\footnote{Here $H(.)$ denotes the binary entropy function.}
\end{theorem}

{\bf Note:} Our assumption that $\dbnd = o(\bl)$ implies that the bounds in Theorem~\ref{thm1} and~\ref{thm2} are $\Omega{\left ( \dbnd\log(\bl) \right )}$.

\section{Main Results}
\label{sec:Result}
All logarithms in this work are assumed to be binary, except where explicitly identified as otherwise (in some cases we explicitly denote the natural logarithm as $\ln(.)$). We define $\Gamma$ as $\ln(\dbnd)/\ln(\bl)$.

\subsection{Upper Bounds on the number of tests required}
The main contributions of this work are explicit computations of the number of tests required to give a desired probability of error via computationally efficient algorithms. Some of these algorithms are essentially new (LP-based decoding algorithms in Theorems~\ref{thm:NCBP-LP}-\ref{thm:CBP-LP}) (though versions of these algorithms were stated with minimal analysis in~\cite{MalS:09,Mal:10,MalM:12}). Others have tighter analysis than previously in the literature (\cm and \ncm) novel analysis (coupon-collector based analysis for \cc, perturbation analysis for the LP-based algorithms). In each case, to the best of our knowledge ours is the first work to explicitly compute the tradeoff between the number of tests required to give a desired probability of error, rather than giving order of magnitude estimates of the number of tests required for a ``reasonable" probability of success. 

First, the \cc algorithm has a novel connection to the Coupon Collector's problem.

\begin{theorem}
\label{thm:CBP}
\cc with error probability at most $\bl^{-\logerror}$ requires no more than $2e\dbnd(1+\logerror)\ln(\bl)$ tests.
\end{theorem}

Next, we consider the ``column matching'' algorithms.

\begin{theorem}
\label{thm4} \label{thm:4} \label{thm:COMP}
\cm with error probability at most $\bl^{-\logerror}$ requires no more than $e\dbnd(1+\logerror)\ln(\bl)$ tests.
\end{theorem}

Translating \cm into the noisy observation scenario is non-trivial. A more careful analysis for the thresholded scheme in \ncm leads to the following result.
We define $\gamma$ as $\frac{\Gamma + \logerror}{1 + \logerror}$ (recall that $\Gamma$ was defined as $\ln(\dbnd)/\ln(\bl)$, and note that $\Gamma$ lies in the interval $[0,1)$ and $\gamma$ in the interval $(\logerror/(\logerror+1),1]$. Let the internal parameters $\threshold$ and $\pmat$ for the remaining algorithms be defined as $\threshold = \frac{1-2\defp}{\defp(1+\gamma^{-1/2})}$ and $\pmat = \frac{1}{\dbnd}$.

\begin{theorem}
\label{thm6} \label{thm:6} \label{thm:NCOMP}
\ncm with error probability at most $\bl^{-\logerror}$ requires no more than $
\frac{16(1+\sqrt{\gamma})^2(1 + \logerror)\ln 2 }{(1-e^{-2})(1-2\defp)^2}\dbnd\log \bl$ tests.
\end{theorem}

Finally, we consider our LP-based algorithms. The main LP-based algorithm, whose analysis dominates the latter half of this work, is \nlpm, which considers the same noisy-measurement scenario that \ncm considers.\footnote{The analysis of the constants in the LP-based algorithms are not optimized (doing so is analytically very cumbersome), but are given to demonstrate the functional dependence on $\logerror$ and $\defp$. 
}

\begin{theorem}
 \label{thm:NCBP-LP}
\nlpm with error probability at most $\bl^{-\logerror}$ requires no more than $\beta_{LP}\dbnd\log \bl$ tests, with $\beta_{LP}$ defined as
{\begin{align*}
\frac{(\logerror + 1 + \Gamma)\ln(2) 2e^2w(w + (1-2\defp)/3)}{(1-2\defp)^2},
\end{align*}}
with $w =   1  + \frac{(1-2\defp)}{\dbnd}$.
\end{theorem}

{
In fact, this algorithm is robust to other types of measurement noise as well. For instance, in the asymmetric noise model, we can make the following statement.

\begin{theorem}
 \label{thm:asym}
In the asymmetric noise model, \nlpm with error probability at most $\bl^{-\logerror}$ requires no more than $\beta_{ASY-LP}\dbnd\log \bl$ tests, with $\beta_{ASY-LP}$ defined as
\begin{align*}
\frac{(\logerror + 1 + \Gamma)\ln(2) 2e^2w(w + (1-\defpn-\defpp)/3)}{(1-\defpn-\defpp)^2},
\end{align*}
with $w =   1  + \frac{(1-\defpn-\defpp)}{\dbnd}$.
\end{theorem}

Similarly, in the activation noise model, we can make the following statement.

\begin{theorem}
 \label{thm:act}
In the activation noise model, \nlpm with error probability at most $\bl^{-\logerror}$ requires no more than $\beta_{ACT-LP}\dbnd\log \bl$ tests, with $\beta_{ACT-LP}$ defined as
\begin{align*}
\frac{(\logerror + 1 + \Gamma)2\ln(2) e^2w\left ( 3w(1+\dil-\defpn)-(1-\dil-\defpn) \right )}{3(1-\dil-\defpn)^2},
\end{align*}
with $w =  1  + \frac{1-\dil-\defpn}{\dbnd}$.
\end{theorem}
}

For the same noiseless observation scenario as CBP, the LP-based decoding algorithm CBP-LP (and which is implied by the stronger analysis (for the noisy observations scenario) for NCBP-LP in Theorem~\ref{thm:NCBP-LP}) has the following performance guarantees. 
\begin{theorem}
\label{thm:CBP-LP}
\lp with error probability at most $\bl^{-\logerror}$ requires no more than $\beta\dbnd\log \bl$ tests, with $\beta$ set as
$$
(\logerror + 1 + \Gamma)\ln(2) 2e^2w'(w' + 1/3),  \mbox{ with } w' =   1  + \frac{1}{\dbnd}.
$$
\end{theorem}

{\bf Note:} Our achievability schemes in Theorems~\ref{thm:COMP}-\ref{thm:CBP-LP} are commensurate (equal up to a constant factor) with the lower bounds in Theorems~\ref{thm:ITLowBnd} and~\ref{thm:ITNoisyLowBnd}. For instance, the bound on the number of tests in Theorem~\ref{thm:NCOMP} differs from the corresponding lower bound in Theorem~\ref{thm:ITNoisyLowBnd}  by a factor that is at most
$12.83(1+\sqrt{\gamma})^2(1 + \logerror)(1-2\defp)^{-2}$, which is a function only of $\logerror$ (which is any positive number for a polynomially decaying probability of error), $\gamma$ (which depends on the scaling between $\dbnd$ and $\bl$ and lies between $\logerror$ and $1$), and $\defp$. However, taking the Taylor series expansion of $1-H(\defp)$ about $1/2$, for all $\defp \in (0,1/2)$, $1-H(\defp)$ is always between $(1-2\defp)^2$ and $(1-2\defp)^2/(2\ln(2))$, taking those extreme values for extreme values of $\defp\in(0,1/2)$. Hence for all values of $\defp$, there is a finite (and explicitly computable) constant-factor gap between our information-theoretic lower bounds, and the upper bounds achieved by our algorithms.

\section{Proof of the performance of algorithms in Theorems~\ref{thm:CBP}-\ref{thm:CBP-LP}}
\label{sec:uprbnd}

\subsection{Coupon collector algorithm}
We first consider \cc, whose analysis is based on a novel use of the Coupon Collector Problem~\cite{Chernoff}.

\noindent {\bf Proof of Theorem~\ref{thm:CBP}}:

The Coupon Collector's Problem (CCP) is a classical problem that considers the following scenario. Suppose there are $n$ types of coupons, each of which is equiprobable. A collector selects coupons (with replacement) until he has a coupon of each type. What is the distribution on his stopping time? It is well-known (\cite{Chernoff}) that the expected stopping time is $\bl\ln \bl + \Theta(\bl)$. Also, reasonable bounds on the tail of the distribution are also known -- for instance, it is known that the probability that the stopping time is more than $\chi \bl \ln \bl$ is at most $\bl^{-\chi + 1}$.

Analogously to the above, we view the group-testing procedure of \cc as a Coupon Collector Problem. Consider the following thought experiment. Suppose we consider any test as a length-$g$ {\it test-vector}\footnote{Note that this test-vector is different from the binary length-$\bl$ vectors that specify tests in the group testing-matrix, though there is indeed a natural bijection between them.} whose entries index the items being tested in that test (repeated entries are allowed in this vector, hence there might be less than $g$ distinct items in this vector). Due to the design of our group-testing procedure in
\cc, the probability that any item occurs in any location of such a vector is uniform and independent. In fact this property (uniformity and independence of the value of each entry of each test) also holds {\it across} tests. Hence, the items in any subsequence of $k$ tests may be viewed as the outcome of a process of selecting a single chain of $gk$ coupons. This is still true even if we restrict ourselves solely to the tests that have a negative outcome. The goal of \cc may now be viewed as the task of collecting {\it all} the {\it non-defective} items. This can be summarized in the following equation

\begin{equation}
\label{eq:1}
\test g\left ( \frac{\bl-\dfct}{\bl} \right )^g \geq (\bl-\dfct)\ln (\bl-\dfct) .
\end{equation}

The left-hand side of Equation (\ref{eq:1}) refers to the expected number of (possibly repeated) items in negative tests (since there are a total of $\test$ tests, each containing $g$ (possibly repeated) items, and the probability of a test being negative equals $((\bl-\dfct)/\bl)^g$). The right-hand side of (\ref{eq:1}) refers to the expected stopping-time of the underlying CCP. We thus optimize (\ref{eq:1}) w.r.t. $g$ to obtain an optimal value of $g$ equaling $1/ \ln (\bl/(\bl-\dfct))$.
However, since the exact value of $\dfct$ is not known, but rather only $\dbnd$, an upper bound on it, we set $g$ to equal $1/ \ln (\bl/(\bl-\dbnd))$.
Taking the appropriate limit of $\bl$ going to infinity, and noting $\dbnd = o(\bl)$, enables us to determine that, in expectation over the testing process and the location of the defective items, (\ref{eq:1}) implies that $\test \geq e\dbnd \ln \bl$.

However, (\ref{eq:1}) only holds in expectation. For us to design a testing procedure for which we can demonstrate that the number of tests decays to zero as $\bl^{-\logerror}$, we need to
modify (\ref{eq:1}) to obtain the corresponding tail bound on $\test$. This takes a bit more work.

The right-hand side is then modified to $\chi (\bl-\dfct) \ln (\bl-\dfct)$. This corresponds to the event that all types of coupons have not been collected if $\chi (\bl-\dfct) \ln (\bl-\dfct)$ total coupons have been collected. The probability of this event is at most $(\bl-\dfct)^{-\chi+1}$).

The left-hand side is multiplied with $(1-\rho)$, where $\rho$ is a design parameter to be specified by the Chernoff bound on the probability that the actual number of items in the negative tests is smaller than $(1-\rho)$ times the expected number. By the Chernoff bound this is at most $\exp \left (-\rho^2T \left ( \frac{\bl-\dfct}{\bl}  \right )^g\right )$. Taking the union bound over these two low-probability events gives us that the probability that
\begin{equation}
\label{eq:2}
(1-\rho)Tg\left ( \frac{\bl-\dfct}{\bl} \right )^g \geq \chi (\bl-\dfct)\ln (\bl-\dfct)
\end{equation}
does {\it not} hold is at most
\begin{equation}
\exp \left (-\rho^2 \test \left(\frac{\bl-\dfct}{\bl}\right)^g \right ) + (\bl-\dfct)^{-\chi + 1}.
\label{motherfucker}
\end{equation}

So, we again optimize for $g$ in (\ref{eq:1}) and substitute $g^{\ast} = 1/\ln \left ( \frac{\bl}{\bl-\dbnd} \right )$ into (\ref{eq:2}). We note that since both $\dbnd$ and $\dfct$ are $o(\bl)$, $\left ( \frac{\bl-\dfct}{\bl}\right )^{g^{\ast}}$ converges to $e^{-1}$. Hence we have, for large $\bl$,
\begin{eqnarray}
	\test & \geq & \frac{\chi}{1 - \rho} \frac{(\bl-\dfct) \ln (\bl-\dfct)}{g^{*} \left(\frac{\bl-\dfct}{\bl}\right)^{g^{*}}} \nonumber \\
	&\approx& \frac{\chi}{1 - \rho} \frac{(\bl-\dfct) \ln (\bl-\dfct)}{\frac{1}{\ln \left(\frac{\bl}{\bl-\dbnd}\right)} e^{-1}} \nonumber \\
	&=& \frac{\chi}{1 - \rho} \frac{(\bl-\dfct) \ln (\bl-\dfct) \ln \left(\frac{\bl}{\bl-\dbnd}\right)}{ e^{-1}}. \label{bb}
\end{eqnarray}

Using the inequality $\ln(1+x) \geq x - x^2/2$ with $x$ as $\dbnd/(\bl-\dbnd)$ simplifies the RHS of (\ref{bb}) to
\begin{equation}
	\test \geq \frac{\chi}{1 - \rho} e \left( \dbnd - \frac{\dbnd^2}{2(\bl-\dfct)} \right) \ln (\bl-\dfct).
\label{bb2}
\end{equation}

Choosing $\test$ to be greater than the bound in (\ref{bb2}) can only reduce the probability of error, hence choosing
\begin{equation}
\test \geq \frac{\chi}{1-\rho}e \dbnd\ln(\bl-\dfct) \nonumber
\label{iloveyou}
\end{equation}
\noindent still implies a probability of error at most as large as in (\ref{motherfucker}).

Choosing $\rho = \frac{1}{2}$, noting that $\dbnd \geq \dfct$, and substituting (\ref{iloveyou}) into (\ref{motherfucker}) implies, for large enough $\dfct$, the probability of error $P_e$ satisfies
\begin{eqnarray}
		 P_e &\leq & e^{-\frac{\delta^2 \chi}{1 - \delta}\dfct \ln (\bl-\dfct)} + (\bl-\dfct)^{-\chi+1} \nonumber \\
		 &=& (\bl-\dfct)^{-\frac{\delta^2}{1-\delta} \chi \dfct} + (\bl-\dfct)^{-\chi+1} \nonumber \\
		 &\leq & 2(\bl-\dfct)^{-\chi+1}. \nonumber
\end{eqnarray}

Taking $2(\bl-\dfct)^{-\chi+1} = \bl^{-\delta}$, we have $\chi = \delta \frac{\log \bl}{\log (\bl-\dfct)} + \frac{1}{\log (\bl-\dfct)} + 1$. For large $\bl$, $\chi $ approaches $ \delta + 1$.

Therefore, the probability of error is at most $n^{-\delta}$, with sufficiently large $n$, the following number of tests suffice to satisfy the probability of error condition stated in the theorem.
\[
	\test \geq 2 (1+\delta) e \dbnd \ln \bl.
\]
\hfill $\blacksquare$

\subsection{Column Matching algorithms}
We next consider Column Matching algorithms. The \cm and \ncm algorithms respectively deal with the noiseless and noisy observation scenarios.

\noindent {\bf Proof of Theorem~\ref{thm:COMP}:}\\
As noted in the discussion on \cm in Section~\ref{subsec:algorithm}, the error-events for the algorithm correspond to false positives, when a column of $\mmat$ corresponding to a non-defective item is ``hidden" by other columns corresponding to defective items. To calculate this probability, recall that each entry of $\mmat$ equals one with probability $\pmat$, i.i.d. Let $j$ index a column of $\mmat$ corresponding to a non-defective item, and let $j_1,\ldots, j_\dfct$ index the columns of $\mmat$ corresponding to defective items. Then the probability that $m_{i,j}$ equals one, and at least one of $m_{i,j_1}, \ldots, m_{i,j_\dfct}$ {\it also} equals one is $\pmat(1-(1-\pmat)^\dfct)$. Hence the probability that the $j$th column is hidden by a column corresponding to a defective item is $\left ( 1-\pmat(1-\pmat)^\dfct \right )^T$. Taking the union bound over all $\bl-\dfct$ non-defective items gives us that the probability of false positives is bounded from above by
\begin{equation}
\label{eq:COMP_false_positive}
P_e = P_e^+ \leq (\bl-\dfct)\left ( 1-\pmat(1-\pmat)^\dfct \right )^T.
\end{equation}
By differentiation, optimizing (\ref{eq:COMP_false_positive}) with respect to $\pmat$ suggests choosing $\pmat$ as $1/\dfct$. However, the precise value of $\dfct$ may not be known, only $\dbnd$, an upper bound on it, might be.
Substituting the value $\pmat = 1/\dbnd$ back into (\ref{eq:COMP_false_positive}), and setting $\test$ as $\beta \dbnd \ln \bl$ gives us
\begin{eqnarray}
P_e & \leq & (\bl-\dfct)\left ( 1-\frac{1}{\dbnd } \left (1-\frac{1}{\dbnd }\right)^{\dfct}\right )^{\beta \dbnd \ln \bl} \nonumber  \\
	          & \leq & (\bl-\dfct)\left ( 1-\frac{1}{\dbnd e} \right )^{\beta \dbnd \ln \bl} \label{eq:dltD}\\
		& \leq & (n-d)e^{-\beta e^{-1}\ln n} \nonumber \\
		& \leq & n^{1 - \beta e^{-1}}.
\end{eqnarray}
Inequality (\ref{eq:dltD}) follows from the previous since $\dfct \leq \dbnd$ by definition, and since $(1-1/x)^x \geq e^{-1}$.
Choosing $\beta = (1+\logerror)e$ thus ensures the required decay in the probability of error.
 Hence choosing $\test$ to be at least $(1+\logerror)e\dbnd\ln \bl$ suffices to prove the theorem. \hfill $\blacksquare$

\vspace{0.2in}

\noindent {\bf Proof of Theorem~\ref{thm:NCOMP}:}\\
Due to the presence of noise leading to both false positive and false negative test outcomes, both false defective estimates and false non-defective estimates may occur in the \ncm algorithm -- the overall probability of error is the sum of the probability of false defective estimates and that of false non-defective estimates. As in the previous algorithm, we set $\pmat=1 / \dbnd$
and $\test= \beta \dbnd \log \bl$. We first calculate the probability of false non-defective estimates by computing the probability that more than the expected number of ones get flipped to zero in the result vector in locations corresponding to ones in the column indexing the defective item. This can be computed as
\begin{eqnarray}
{P}_e^- & = & \bigcup_{i=1}^{\dfct} P\left ( |{\cal T}_j|=t \right )\Pr\left ( |{\cal S}_j|< |{\cal T}_j|(1-\defp(1+\threshold )) \right )
\nonumber \\
& \stackrel{}{\leq} & \dfct\sum_{t=0}^{\test}{\test \choose t}\pmat^t(1-\pmat)^{\test-t}\label{justification:na}
\\ && \: \sum_{r=t-t(1-\defp(1+\threshold ))}^{t}{t \choose r}\defp^r(1-\defp)^{t-r}
\nonumber \\
& \stackrel{\label{justification:nb}}{\leq} & \dfct\sum_{t=0}^{\test}{\test \choose t}\pmat^t(1-\pmat)^{\test-t}e^{-2t(\defp\threshold)^2}
\\
& \stackrel{\label{justification:nc}}{=} & \dfct \left ( 1-\pmat +\pmat e^{-2(q\threshold)^2} \right )^\test
\\
& \stackrel{\label{justification:nd}}{=} & \dfct \left ( 1-\frac{1}{\dbnd }+\frac{1}{\dbnd}e^{-2(\defp\threshold)^2} \right )^{\beta \dbnd \log \bl}
\\
& \stackrel{\label{justification:ne}}{\leq} & \dfct \exp \left [ {- \beta \log \bl \left ( 1-e^{-2(q\threshold)^2 } \right )} \right ]
\\
& \stackrel{}{\leq} & \dfct \exp \left [ {- \beta \log \bl (1-e^{-2})(\defp \threshold)^2} \right ]\label{justification:nf}
\end{eqnarray}
Here, as in Section~\ref{subsec:algorithm}, ${\cal T}_j$ denotes the locations of ones in the $j$th column of $\mmat$. Inequality (\ref{justification:na}) follows from the union bound over the possible errors for each of the defective items, with the first summation accounting for the different possible sizes of ${\cal T}_j$, and the second summation accounting for the error events corresponding to the number of one-to-zero flips exceeding the threshold chosen by the algorithm. Inequality (\ref{justification:nb}) follows from the Chernoff bound. Equality (\ref{justification:nc}) comes from the binomial theorem. Equality (\ref{justification:nd}) comes from substituting in the values of $\pmat$ and $\test$. Inequality (\ref{justification:ne}) follows from the leading terms of the Taylor series of the exponential function. Inequality (\ref{justification:nf}) follows from bounding the concave function $1-e^{-2x}$ by the linear function $(1-e^{-2})x$ for $x > 0$.

The requirement that the probability of false non-defective estimates $P_e^-$ to be at most $\bl^{-\logerror}$ implies that $\beta^{-}$ (the bound on $\beta$ due to this restriction) satisfies
\begin{eqnarray}
& \ln \left ( \dfct \exp \left [ {-\beta (1-e^{-2})(\defp\threshold)^2 \log \bl}\right ]  \right ) \mbox{ } < & -\logerror \ln \bl
\nonumber \\
\Rightarrow & \ln \dfct-\frac{ \beta (1-e^{-2})(\defp \threshold)^2}{\ln 2}\ln \bl \mbox{ } < & -\logerror \ln \bl
\nonumber \\
\Rightarrow&  \beta^- >  \frac{\left ( \frac{\ln \dfct }{\ln \bl} + \logerror \right ) \ln 2}{
(1-e^{-2})(\defp \threshold)^2}  & \label{beta_n}
\end{eqnarray}


We now focus on the probability of false defective estimates. In the noiseless \cm algorithm, the only way a false defective estimate could occur was if all the ones in a column are hidden by ones in columns corresponding to defective items. In the \ncm algorithm this still happens, but in addition noise could also lead to a similar masking effect. That is, even in the $1$ locations of a column corresponding to a non-defective not hidden by other columns corresponding to defective items, measurement noise may flip enough zeroes to ones so that the decoding threshold is exceeded, and the decoder hence incorrectly declares that particular item to be defective. See Figure~\ref{fig:ncomp}(a) for an example.

Hence we define a new quantity $a$, which denotes the probability for any $(i,j)$th location in $\mmat$ that a $1$ in that location is ``hidden by other columns {\it or} by noise". It equals
\begin{eqnarray}
a & = &1-[(1-\defp)(1-\pmat)^\dfct+\defp(1-(1-\pmat)^\dfct)]
\nonumber \\
& = & \left (1-\defp- \left (1-\frac{1 }{\dbnd} \right )^\dfct(1-2\defp) \right )
\end{eqnarray}
We set $\dbnd \geq 2$ (the case $\dbnd = 1$ can be handled separately by the same analysis, but setting $\pmat = 1/(\dbnd+1) = 1/2$ rather than $1/\dbnd = 1$), and note that by definition $\dfct \leq \dbnd$.
We then bound $a$ from above as
\begin{IEEEeqnarray}{rCl}
\max_{\dbnd \geq 2, \dfct \leq \dbnd} a 
& =  & \max_{\dbnd \geq 2, \dfct \leq \dbnd}  \left (1-\defp- \left (1-\frac{1 }{\dbnd} \right )^\dfct(1-2\defp) \right )
\nonumber \\
& =  & 1-\defp - (1-2\defp) \min_{\dbnd \geq 2, \dfct \leq \dbnd}  \left(\!\!\left (1-\frac{1 }{\dbnd}\right )^\dfct \right )
\label{eq:Dd} \\
& =  & 1-\defp - (1-2\defp) \min_{\dbnd \geq 2}  \left ( \left (1-\frac{1 }{\dbnd}\right )^\dbnd \right )
\label{eq:Dgt2} \\
& =  & (1-\defp) - (1-2\defp)/4.
\label{false positive}
\end{IEEEeqnarray}
Equation (\ref{false positive}) follows from the observations that (\ref{eq:Dd}) is optimized when $\dfct = \dbnd$ and (\ref{eq:Dgt2}) is optimized when $\dbnd = 2$.
The probability of false defective estimates is then computed in a similar manner to that of false non-defective estimates as in (\ref{justification:na})--(\ref{justification:nf}).
\begin{IEEEeqnarray}{rCl}
P_e^+ & = & \bigcup_{i=1}^{\bl-\dfct} P\left ( |{\cal T}_j|=t \right )P\left ( |{\cal S}_j|\geq |{\cal T}_j|(1-\defp(1+\threshold )) \right )
\nonumber \\
& \leq & (\bl-\dfct)\sum_{t=0}^{T}{T \choose t}\pmat^t(1-\pmat)^{\test-t} \nonumber \\ && \: \sum_{r=t(1-\defp(1+\threshold ))}^{t}{t \choose r}a^r(1-a)^{t-r}
\nonumber \\
& \leq  & (\bl-\dfct)\left ( 1-\pmat+\pmat e^{-2((1-\defp(1+\threshold ))-a)^2} \right )^\test
\label{justification:pa} \\
& \leq  & (\bl-\dfct)  \left ( 1-\pmat+\pmat e^{-2 \left ((1-2\defp)/4 - \defp\threshold \right )^2} \right )^\test
\label{justification:pb} \\
& \leq & (\bl-\dfct) \exp  \left [ - ( 1-e^{-2 \left ((1-2\defp)/4- \defp\threshold \right )^2}) \beta\log \bl \right ]
\label{justification:pc}\\
& \leq & \!(\bl-\dfct) \exp  \!\left [ - \beta\log \bl( 1-e^{-2})\! \left (\!\frac{(1-2\defp)}{4} -\defp\threshold\right )^2  \right ]
\label{justification:pd}
\end{IEEEeqnarray}
Note that for the Chernoff bound to be applicable in (\ref{justification:pa}), $1-\defp(1+\threshold)>a$, which implies that $\threshold < (1-2\defp)/(4\defp)$. Equation (\ref{justification:pb}) follows from substituting the bound derived on $a$ in (\ref{false positive}) into (\ref{justification:pa}), and  (\ref{justification:pc}) follows by substituting $\pmat = 1/\dbnd$ into the previous equation.
Inequality (\ref{justification:pd}) follows from bounding the concave function $1-e^{-2x}$ by the linear function $(1-e^{-2})x$ for $x > 0$.

The requirement that the probability of false defective estimates $P_e^+$ be at most $n^{-\logerror}$ implies that $\beta^{+}$ (the bound on $\beta$ due to this restriction) be at least 
\begin{equation}
\beta ^+> \frac{\left ( \frac{\ln (\bl-\dfct)}{\ln n} + \logerror \right ) \ln 2}{(1-e^{-2})((1-2\defp)/4- \defp\threshold)^2}.
\label{beta_p}
\end{equation}



%

Note that $\beta$ must be at least as large as $\max\{\beta^{-},\beta^{+}\}$ so that both (\ref{beta_n}) and (\ref{beta_p}) are satisfied.

When the threshold in the \ncm algorithm is high ({\it i.e.,} $\threshold$ is small) then the probability of false negatives increases; conversely, the threshold being low ($\threshold$ being large) increases the probability of false defective estimates. Algebraically, this expresses as the condition that $\threshold > 0$ (else the probability of false non-defective estimates is significant), and conversely to the condition that $1-\defp(1+\threshold) > a$ (so that the Chernoff bound can be used in (\ref{justification:pa})) -- combined with (\ref{false positive}) this implies that $\threshold \leq (1-2\defp)/4\defp$. Each of (\ref{beta_n}) and (\ref{beta_p}) as a function of $\threshold$ is a reciprocal of a parabola, with a pole at the corresponding extremal value of $\threshold$. Furthermore, $\beta^{-}$ is strictly increasing and $\beta^{+}$ is strictly decreasing in the region of valid $\threshold$ in $(0,(1-2\defp)/(4\defp))$. Hence the corresponding curves on the right-hand sides of (\ref{beta_n}) and (\ref{beta_p}) intersect within the region of valid $\threshold$, 
and a good choice for $\beta$ is at the $\threshold$ where these two curves intersect. To find this $\beta$, we make another simplifying substitution. Let $\gamma$ be defined as\footnote{Note that we replace $\dfct$ with $\dbnd$ in (\ref{beta_n})
and (\ref{beta_p}) since this still leaves the inequalities true, but
now converts $\beta^-$ and $\beta^+$ into quantities that are
independent of $\dfct$. This is necessary since $\beta^-$ and
$\beta^+$ are (internal) code design parameters, and as such must be
independent of the true value of $\dfct$ (which might be unknown) and
may depend only on the upper bound $\dbnd$.}
\begin{equation}
\gamma = \frac{\ln \dbnd + \logerror \ln \bl}{\ln (\bl-\dfct) + \logerror \ln \bl}.
\label{logerror}
\end{equation}
and $\Gamma$ as
\begin{equation}
\Gamma = \frac{\ln \dbnd}{\ln \bl}.
\nonumber
\end{equation}
Hence $\gamma$ for large $\bl$ essentially equals
\begin{equation}
\frac{\Gamma + \logerror }{1 + \logerror}.
\label{eq:Logerror}
\end{equation}
(Note that since $\Gamma$ lies in the interval $[0,1)$, hence $\gamma$ lies in the interval $[\logerror/(\logerror+1), 1)$.) 

Then equating the RHS of (\ref{beta_n}) and (\ref{beta_p}) implies that the optimal $\threshold^\ast$ satisfies
\begin{IEEEeqnarray}{rCl}
\frac{\ln 2}{ (1-e^{-2})((1-2\defp)/4-\defp\threshold^\ast )^2} & = & \frac{\gamma \ln 2}{ (1-e^{-2})(\defp\threshold^\ast )^2}
\label{eq:betas}
\end{IEEEeqnarray}
Simplifying (\ref{eq:betas}) gives us that
\begin{equation}
\label{epsilon_star}
\threshold^* = \frac{1-2\defp}{4\defp(1+\gamma^{-1/2})}.
\end{equation}

Substituting these values of $\gamma$ and $\threshold$ into (\ref{beta_n}) gives us the explicit bound for large $\bl$
\begin{equation}
\label{beta_star}
\beta^\ast = \frac{16(1+\gamma^{-0.5})^2(\Gamma + \logerror)\ln 2 }{(1-e^{-2})(1-2\defp)^2}.
\end{equation}
Using (\ref{eq:Logerror}) to simplify (\ref{beta_star}) gives
\begin{equation}
\beta^\ast = \frac{16(1+\sqrt{\gamma})^2(1 + \logerror)\ln 2 }{(1-e^{-2})(1-2\defp)^2}
\approx \frac{12.83(1+\sqrt{\gamma})^2(1 + \logerror)}{(1-2\defp)^2}.
\nonumber
\end{equation}
\hfill $\blacksquare$

\subsection{LP-decoding algorithms}

For the LP-based algorithms, we first prove Theorem~\ref{thm:NCBP-LP}, and then derive Theorems~\ref{thm:asym},~\ref{thm:act} and~\ref{thm:CBP-LP} 
 as direct corollaries.

\noindent {\bf Proof of Theorem~\ref{thm:NCBP-LP}:}\\
\noindent
{\bf \underline{Proof sketch:}}
For the purpose of exposition we break the main ideas into four steps below. 

\noindent
{\bf (1)}\,\,\underline{Finite set of Perturbation Vectors:}  For the known $\dfct$ case we define a {\it finite} set $\Prt'$ (that depends on the true $\bx$)
containing so-called ``perturbation vectors".\footnote{This set is defined just for the purpose of this proof -- the encoder/decoder do not need to know this set.}
We demonstrate in Claim~\ref{clm:xset} that any $\bar{\bx}$ in the feasible set of the constraint set of \nlpm can be written as the true $\bx$ plus a {\it non-negative linear combination} of  perturbation vectors from this set. The physical intuition behind the proof is that the vectors from $\Prt'$ correspond to a ``mass-conserving" perturbation of $\bx$.
The property of non-negativity of the linear combinations arises from a physical argument demonstrating that there is a path from $\bx$ to any point in the feasible set using these perturbations, over which one never has to ``back-track". The linear combination property is important, since this enables us to characterize the directions in which a vector can be perturbed from $\bx$ to another vector that satisfies the constraints of \nlpm, in a ``finite" manner (instead of having to consider the uncountably infinite number of directions that $\bx$ could be perturbed to). The non-negativity of the linear combination is also crucial since, as we explain below, this property ensures that the objective function of the LP can only increase when perturbed in a convex combination of the directions in $\Prt'$.

\noindent
{\bf (2)}\,\,\underline{Expected Perturbation Cost:}
The heart of our argument then lies in Claims~\ref{clm:val_del} and~\ref{clm:ecstdel}, where we characterize via an exhaustive case-analysis the expected change (over randomness in the matrix $\mmat$ and noise $\nu$) in the value of each slack variable $\cost_i$ when $\bx$ is perturbed to some ${\bx'}$ by a vector in $\Prt'$.
In particular, we demonstrate that for each such {\it individual} perturbation vector, the expected change in the value of each slack variable $\cost_i$ is {\it strictly} positive with high probability.\footnote{Actually, for each fixed feasible $\bx$, due to equality (\ref{eq:ncbp1}) there may be a range of $\cost_i$ variables such that $(\bx,\mathbf{\cost})$ are simultaneously feasible in \nlpm. 
Here and in the rest of this paper we abuse notation by using $\cost_i$ to denote specifically those variables that meet all constraints in \nlpm with equality, and hence correspond to the smallest possible value of the objective function of \nlpm for that fixed $\bx$.}
The actual proof follows from a case-analysis similar to the one performed in the example in Table~\ref{tab:allfour}.

\noindent
{\bf (3)}\,\,\underline{Concentration \& Union Bounding:}
Next, in Claim~\ref{clm:eLPcst} with slightly careful use of standard concentration inequalities (specifically, we need to use both the additive and multiplicative forms of the Chernoff bound, reprised in Claim~\ref{clm:sanov}) we show that the probability distributions derived in Claim~\ref{clm:ecstdel} are sufficiently concentrated.
We then take the union bound over all vectors in $\Prt'$ (in fact, there are a total of $\dfct(\bl-\dfct)$ such vectors in $\Prt'$) and show that with sufficiently high probability the expected change in the value of the objective function (which equals the weighted sum of the changes in the values of the slack variables $\cost_i$) for {\it each} perturbation vector in $\Prt'$ is also {\it strictly} positive.


\noindent
{\bf (4)}\,\,\underline{Generalization Based on Convexity:}
We finally note in Claim~\ref{clm:eLPcst} that the set of feasible $(\bar{\bx},\cost)$ of \nlpm for a fixed value of $\dfct$ forms a convex set. Hence if the value of the LP objective function strictly increases along every direction in $\Prt'$, then in fact the value of the LP objective function strictly increases when the true $\bx$ is perturbed in {\it any} direction (since, as noted before, any vector in the feasible set can be written as $\bx$ plus a non-negative linear combination of vectors in $\Phi'$). Hence the true $\bx$ must be the solution to \nlpm. This completes our proof.

\noindent
{\bf \,\,\underline{Proof details:}}
We now proceed by proving the sequence of claims that when strung together formalize the above argument.

Without loss of generality, let $\bx$ be the vector with $1$s in the first $\dfct$ locations, and $0$s in the last $\bl-\dfct$ locations.\footnote{As can be verified, our analysis is agnostic to the actual choice of $\bx$, as long as it is a vector in $\{0,1\}^{\bl}$ of weight any $\dfct \leq \dbnd$.}
Choose $\Prt'=\{\prt'\}_{k'=1}^{\dfct(\bl-\dfct)}$ as the set of $\dfct(\bl-\dfct)$ vectors with a single $-1$ in the support of $\bx$, a single $1$ outside the support of $\bx$, and zeroes everywhere else. For instance, the first $\prt'$ in the set equals $(-1,0,\ldots,0,1,0,\ldots,0)$, where the $1$ is in the $(\dfct+1)$th location. 

Then, Claim~\ref{clm:xset} below gives a ``nice" characterization of the set of $\bar{\bx}$ in the feasible set of \nlpm.

\begin{claim}
Any vector $\bar{\bx}$ with the same weight as $\bx$ ({\it i.e.}, if $\brd = \dfct$) that satisfies the constraints (\ref{eq:ncbp1}--\ref{eq:ncbp5}) in NCBP-LP can be written as
\begin{equation}
\bar{\bx} = \bx + \sum_{k'=1}^{\dfct}\sum_{k''=\dfct+1}^{\bl}c_{k',k''}\prt'_{k',k''}, \mbox{ with all } c_{k',k''} \geq 0.
\label{eq:prtvec}
\end{equation}
\label{clm:xset}
\end{claim}
\noindent {\it Proof of Claim~\ref{clm:xset}:} The intuition behind this claim is by ``conservation of mass", so to speak. A good analogy is the following physical process.

Imagine that $\bx$ corresponds to $\bl$ bottles of water each with capacity one litre, with the first $\dfct$ bottles full, and the remaining empty. Imagine $\bar{\bx}$ as another state of these bottles with $\bar{\dfct} = \dfct$ litres of water.
For each bottle $j$ among the first $\dfct$ bottles that has more water remaining than in the corresponding bottle in the final state $\bar{x}_i$, we use its water to increase the water level of bottles among the last $\bl-\dfct$ bottles (taking care not to overshoot, {\it i.e.,} not to put more than the desired water level $\bar{x}_i$ in any such bottle). The fact that this is doable follows from conservation of mass. This corresponds to using non-negative linear combinations of perturbation vectors from the set $\Prt'$ (non-negativity arises from the fact the we took care not to overshoot).

In fact, amongst various ways to do this, the following specific choices of $c_{k',k''}$ also work. Define $C$ as $\sum_{k''=\dfct+1}^\bl \bar{x}_{k''}$, {\it i.e.,} the amount of water transferred to the empty bottles. We then set $c_{k',k''} = (1-\bar{x}_{k'})\bar{x}_{k''}/C$. That is, first consider the amount of water transferred out of the $k'$ bottle -- this equals $(x_{k'}-\bar{x}_{k'})$, which equals $1-\bar{x}_{k'}$ since $x_{k'} = 1$ by assumption. We apportion this water in proportion to the water required in each of the $k''$th bottles.
\hfill $\Box$

Next, Claim~\ref{clm:ecstdel} below computes the expected change in the value of the slack variable $\cost_i$ as $\bx$ is perturbed by $\prt'$. A small example in Table~\ref{tab:allfour}
(with $\bl = 3$, $\dfct = 2$) demonstrates the calculations in the proof of Claim~\ref{clm:ecstdel} explicitly.

For any fixed $\bx \in \{0,1\}^{\bl}$ of weight $\dfct \leq \dbnd$, let ${\bx}' = \bx + \prt'$. Over the randomness in $\mmatr_i$ and the noise $\nu_i$ generating the testing outcome $\hat{y}_i = y_i+\nu_i$, we define the {\it cost perturbation} random variables
\begin{equation}
\begin{array}{cc}
\Delta_{0,i}'  =  (\cost_i(\bx')-\cost_i({\bx})), 
\mbox{ conditioned on } \hat{y}_i=0.
\end{array}\label{eq:del_defn}
\end{equation}

\begin{claim}
The cost perturbation random variables defined in (\ref{eq:del_defn}) all take values only in $\{-1,0,1\}$.
\label{clm:val_del}
\end{claim}
\noindent {\it Proof of Claim~\ref{clm:val_del}:}
We first analyze the case when if $\hat{y}_i=0$.
By (\ref{eq:ncbp1}), $\cost_i(\bx) = \mmatr_i . \bx$. Hence 
$\Delta'_{0,i} = \cost_i(\bx')- \cost_i(\bx)= \mmatr_i . (\bx'-\bx) = \mmatr_i . \prt'$. But 
$\prt'$ has exactly one component equaling $-1$ and one equaling $1$. By definition, $\mmatr_i$, is a $0/1$ vector.
Hence 
$\Delta'_{0,i}$ takes values only in $\{-1,0,1\}$.

\hfill $\Box$

The next claim forms the heart of our proof. It does an exhaustive\footnote{And exhausting!} case analysis that computes the probabilities that the cost perturbation random variables take values $1$ or $-1$ (the case that they equal zero can be derived from these calculations in a straightforward manner too, but since these values turn out not to matter for our analysis, we omit these details).

We define the {\it expected objective value perturbation} $\Delta^\test$ as
\begin{align}
\Delta^\test = \sum_{i=1}^\test \left [1(\Delta'_{0,i}=1)-1(\Delta'_{0,i}=-1)\right ],
\label{eq:ovalp}
\end{align}
and the {\it number of perturbed noise variables} $(\#\Delta)^\test$ as
\begin{align}
(\#\Delta)^\test = \sum_{i=1}^\test \left [1(\Delta'_{0,i}=1)+1(\Delta'_{0,i}=-1)\right ].
\label{eq:ovalp2}
\end{align}
\noindent Here $1(.)$ is denotes the indicator function of an event.

\begin{claim}
\begin{align}
& P(\Delta'_{0,i}=1)  =  \pmat(1-\pmat)\left [ (1-2\defp)(1-\pmat)^{\dfct-1}+\defp \right ], \label{eq:costch2a} \\
& P(\Delta'_{0,i}=-1) =  \pmat(1-\pmat)\defp. \nonumber
\end{align}
Hence the expected value of the objective value perturbation $\Delta^\test$ equals $\test\pmat(1-\pmat)^{\dfct}(1-2\defp)$, and the expected value of the number of perturbed variables $(\#\Delta)^\test$ equals $\test\pmat \left [(1-\pmat)^{\dfct}(1-2\defp)+2(1-\pmat)\defp  \right ]$.
\label{clm:ecstdel}
\end{claim}
\noindent {\it Proof of Claim~\ref{clm:ecstdel}:} The proof follows from a case-analysis similar to the one performed in the example in Table~\ref{tab:allfour}. The reader is strongly encouraged to read that example before looking at the following case analysis, which can appear quite intricate.

\begin{table*}
\hspace{0in}
\centering
\begin{tabular}{ | c | c | c | c | c | c || c || c | c | }

\cline{7-8}
\multicolumn{6}{c}{} & \multicolumn{1}{|c||}{$\bx  $} & \multicolumn{1}{c|}{${\bx}' = \bx+\prt' $}  &\multicolumn{1}{c}{}\\

 \multicolumn{6}{c}{} & \multicolumn{1}{|c||}{$ \mathbf{(1,1,0)}$} &
\multicolumn{1}{c|}{$  \mathbf{(0,1,1)}$}  & \multicolumn{1}{c}{}\\\hline

1. & 2. & 3. & 4. & 5. & 6. & 7. & 8(a). & 8(b). \\

$\hat{\by_i}$ & $\cost(\bx)$  & $\by_i$ & $\mmatr_i$  & $P(\hat{\by_i},\mmatr_i|\bx)$ & $\cost_i(\mmatr_i,\bx)$ & $\cost_i(\mmatr_i,\bx)$ &  $\cost_i({\mmatr_i,\bx}')$ &  E($\mmatr_i,\Delta'_i$)
\\ \hline \hline









\multirow{8}{*}{0} & \multirow{8}{*}{$\mmatr_i.\bx$} & \multirow{2}{*}{0} & $\mathbf{(0,0,0)}$ & $(1-\defp) (1-\pmat)^3$ &  0 & 0 &  0 & 0 \\ \cline{4-9}

& &  & $\mathbf{(0,0,1)}$ & $(1-\defp) (1-\pmat)^2p$ & $x_3$ & 0 &  1 \cellcolor{green} & $(1
- \defp) (1-\pmat)^2p$ \\  \cline{3-9}

& & \multirow{6}{*}{1} & $\mathbf{(0,1,0)}$ & $\defp (1-\pmat)^2p$ & $x_2$ & 1 &  1 & 0 \\  \cline{4-9}

& &  & $\mathbf{(0,1,1)}$ & $\defp (1-\pmat)p^2$& $x_2 +x_3$ & 1 &  2 \cellcolor{green} & $\defp(1-\pmat)p^2$ \\ \cline{4-9}

& &  & $\mathbf{(1,0,0)}$ & $\defp (1-\pmat)^2p$ & $x_1$ & 1 & 0 \cellcolor{red} & $-\defp(1-\pmat)^2p$\\ \cline{4-9}

& &  & $\mathbf{(1,0,1)}$ & $\defp (1-\pmat)p^2$ & $x_1 + x_3$ & 1 &  0 & 0 \\ \cline{4-9}

& &  & $\mathbf{(1,1,0)}$ & $\defp (1-\pmat)p^2$ & $x_1 + x_2$ & 2 &  1\cellcolor{red} & $-\defp(1-\pmat)p^2$\\ \cline{4-9}

& &  & $\mathbf{(1,1,1)}$ & $\defp p^3$&  $x_1 + x_2 + x_3$ & 2 &  2 & 0 \\ \hline \hline

\multicolumn{8}{c}{} & \multicolumn{1}{|c|}{\textcolor{blue}{$(1- 2\defp) (1-\pmat)^2p$}}  \\
\cline{9-9}
\end{tabular}
\vspace{0.1in}
\caption{Suppose
$\bx = (1,1,0)$. Choose some ${\bx}' \neq \bx$ (in this example, ${\bx}' = \bx + \prt'$, where $\prt'=(-1,0,1)$ is a {\it perturbation vector}). This example analyzes the expectation (over the randomness in the particular row $\mmatr_i$ of the measurement matrix $\mmat$) of the difference in value of the corresponding slack variables $\cost_i(\bx)$ and $\cost_i({\bx}')$ in column $8(b)$. To compute these, we consider the columns of the table above sequentially from left to right. Column $1$ considers the two possible values of the observed vector $\hat{y}_i$. Column $2$ gives the corresponding values of the slack variables corresponding to the $i$th test, as returned by the constraints (\ref{eq:ncbp1}) of \nlpm -- 
Column $3$ indexes the possibilities of the (noiseless) test outcomes $y_i$, and column $4$ enumerates possible values for $\mmatr_i$, the $i$-th row of $\mmat$, that could have generated the values of $y_i$ in column $3$, given that $\bx  = (1,1,0)$. Column $5$ computes the probability of
a particular observation $\hat{y}_i$ and a row $\mmatr_i$, given that the noiseless output $y_i$ equaled a particular value. Column $6$ computes the function in column $2$, given that $\mmatr_i$ equals the value given in Column $4$. Columns $7$ and $8(a)$ respectively explicitly compute the value of the function in column $6$ for the vectors $\bx$ and ${\bx}'$ -- the red entries in column $8(a)$ index those locations where $\cost({\bx}')$, the slack variable for the perturbed vector, is less than $\cost(\bx)$, and the green cells indicate those locations where the situation is reverse. Column $8(b)$ then computes the product of column $5$ with the difference of the entries in column $7$ from those of column $8(a)$, {\it i.e.}, the expected change in the value of the slack variable $\cost_i(.)$. The value $(1-2\defp)(1-\pmat)^2\pmat$ in blue at the bottom represents the expected change (averaged over all possible tuples $(y_i,\mmatr_i,\hat{y}_i)$).
}
\label{tab:allfour}
\end{table*}

\begin{itemize}

\item Equation (\ref{eq:costch2a}) analyzes $\Delta'_{0,i}=-1$, the expected change in $\cost_i$ if $\bx$ is perturbed by a vector from $\Prt'$, and $\hat{y}_i=0$. By (\ref{eq:ncbp1})
 $\cost_i(\bx) = \mmatr_i . \bx$, hence $\Delta_0' = \cost_i(\bx')- \cost_i(\bx)= \mmatr_i . \prt'$.
 
 We first analyze the case when $\Delta'_{0,i}=-1$. This occurs only when $\mmatr_i = 1$ where $\prt' = -1$ (hence there is non-zero intersection with the support of $\bx$ and so $y_i = 1$), and further that $\mmatr_i$ equals $0$ in the location where $\prt' = 1$. Thus, only $2$ indices of $\mmatr_i$ matter for this scenario. These scenarios occurs with probability $\defp(1-\pmat)\pmat$.

Analogously, the only scenarios when $\Delta'_{0,i}=1$ occur when $\mmatr_i$ equals $1$ in the location where $\prt' = 1$ and $\mmatr_i$ equals $0$ in the location where $\prt' = -1$. This can happen in two ways. It could be that if the support of $\mmatr_i$ is entirely outside the support of $\bx$ (hence $y_i=0$), and further that $\mmatr_i$ equals $1$ in the location where $\prt' = 1$ (this happens with probability $(1-\defp)(1-\pmat)^\dfct\pmat$). Or, it could be that $\mmatr_i$ equals $0$ where $\prt'=-1$, $\mmatr_i$ equals $1$ in at least one of the other $(\dfct-1)$ locations in the support of $\bx$ (which ensures that $y_i=1$), and further that $\mmatr_i$ equals $1$ where $\prt'=1$ (the probability of such a scenario is
$\defp(1-\pmat)(1-(1-\pmat)^{\dfct-1})\pmat$).
Adding together the probabilities corresponding to these two scenarios gives the desired result.
\end{itemize}

Substituting the four terms obtained in (\ref{eq:costch2a}) into (\ref{eq:ovalp}) and (\ref{eq:ovalp2}) gives the required result for $\Delta^\test$ and $(\#\Delta)^\test$.
\hfill $\Box$

We now recall Bernstein's inequality, a classical concentration inequality that we use to make a statement about the concentration about the expectation of the objective function value of \nlpm.

\begin{claim}[Bernstein inequality~\cite{Chernoff}]
Let $\{W_i\}_{i=1}^\test$ be a sequence of independent zero-mean random variables. Suppose $|W_i| \leq w$ for all $i$. Then for all negative $\sigma$,
\begin{eqnarray}
P \left (  \sum_{i=1}^\test {W_i}  < \sigma \right ) & \leq & \exp \left (-\frac{\sigma^2/2}{\sum_i^\test EW_i^2 - w\sigma/3} \right ). \label{eq:sanov}
\end{eqnarray}
\label{clm:sanov}
\end{claim}

Next, Claim~\ref{clm:eLPcst} below demonstrates that if $\bx$ is perturbed in any direction from the set $\Prt'$, as long as it remains within the feasible set for \nlpm, with high probability over $\mmat$ and the noise vector $\nu$, the value of the objective function of \nlpm increases. 

\begin{claim}
Choose $\test$ as $\beta_{LP}\dbnd\log(\bl)$, with
$\beta_{LP}$ as given in Theorem~\ref{thm:NCBP-LP}.
Then with probability no more than $\bl^{-\logerror}$ the value of the objective function of \nlpm has a unique optimum at $\hat{\bx} = \bx$.
\label{clm:eLPcst}
\end{claim}
\noindent {\it Proof of Claim~\ref{clm:eLPcst}:}
Since each of the $\mmatr_i$ vectors and the noise vector $\mathbf{\nu}$ are all chosen independently, the random variables corresponding to each type of cost perturbation variable in (\ref{eq:del_defn}) are distributed i.i.d. according to (\ref{eq:costch2a}).

We now use Claim~\ref{clm:sanov} to concentrate around the expected value given in Claim~\ref{clm:ecstdel}. In particular, for each $i \in \{1,\ldots,\test\}$ we set 
\begin{align}
W_i = (\cost_i(\bx')-\cost_i(\bx)) - E(\cost_i(\bx')-\cost_i(\bx)),
\label{eq:wi}
\end{align} 
\noindent hence each $W_i$ is zero-mean. 

Hence $w$, the maximum value of $|W_i|$, can be bounded as
\begin{align}
w & =  |\cost_i(\bx')-\cost_i(\bx) - E(\cost_i(\bx')-\cost_i(\bx))|\nonumber\\
& \leq  |\cost_i(\bx')-\cost_i(\bx)|  +  |E(\cost_i(\bx')-\cost_i(\bx))|\nonumber \\
& =  1  + \pmat(1-\pmat)^\dfct(1-2\defp)\label{eq:wiub0}\\
& <  1  + (1-2\defp)/\dbnd \label{eq:wiub}.
\end{align} 
\noindent Here (\ref{eq:wiub0}) follows by Claim~\ref{clm:val_del} which demonstrates that $\cost(\bx')-\cost(\bx)$ takes values only in $\{-1,0,1\}$, and from the computation of $E(\cost_i(\bx')-\cost_i(\bx))$ in the last part of Claim~\ref{clm:ecstdel}. Equation (\ref{eq:wiub}) uses the definition of $\pmat$ as $1/\dbnd$ and the bound that $(1-\pmat)^\dfct < 1$. \footnote{As always, the case of $\dbnd = 1$ is handled separately by choosing $\pmat = 1/2$.}

Next, we bound $\sum_{i=1}^{\test}EW_i^2$ from above by
\begin{align}
\sum_{i=1}^{\test}EW_i^2 & \leq  w^2(\#\Delta)^\test \nonumber\\
& =  2w^2\test\pmat \left [(1-\pmat)^{\dfct}(1-2\defp)+2(1-\pmat)\defp  \right ] \label{eq:sumewi2}\\
& <  2w^2\beta_{LP}\log(\bl) \label{eq:sumewi22}.
\end{align} 
\noindent Here (\ref{eq:sumewi2}) follows from Claim~\ref{clm:ecstdel}, and (\ref{eq:sumewi22}) by using $\pmat = 1/\dbnd$,  $\test = \beta_{LP}\dbnd\log(\bl)$, and $(1-\pmat)^\dfct < 1$.

We then set 
\begin{align}
\sigma = -\test E(\cost_i(\bx')-\cost_i(\bx)) = -\beta_{LP}\log(\bl)(1-\pmat)^\dfct(1-2\defp),
\label{eq:sigma}
\end{align}
\noindent which is negative as required. We use $\pmat = 1/\dbnd$, $\test = \beta_{LP}\dbnd\log(\bl)$ and $1/e < (1-1/\dbnd)^\dfct < 1$ to give us that 
\begin{align}
\beta_{LP}\log(\bl)(1-2\defp)/e < |\sigma| <\beta_{LP}\log(\bl)(1-2\defp).
\label{eq:sigmabnd}
\end{align}
\noindent (For the numerator of the exponent of Claim~\ref{clm:sanov} we shall use the lower bound, and for the denominator the upper bound.)

Substituting the value of $\sigma$ from (\ref{eq:sigma}), $W_i$ from (\ref{eq:wi}), noting that $\sigma$ is negative, using Claim~\ref{clm:sanov},
and using bounds for $w$, $\sum_{i=1}^{\test}EW_i^2$ and $\sigma$ respectively from (\ref{eq:wiub}), (\ref{eq:sumewi22}) and (\ref{eq:sigmabnd})
gives us that
\begin{align}
P & \left (  \sum_{i=1}^\test \left ( \cost_i(\bx')-\cost_i(\bx) \right )  < 0 \right ) \\
  & =  P \left (  \sum_{i=1}^\test \left ( (\cost_i(\bx')-\cost_i(\bx)) - E(\cost_i(\bx')-\cost_i(\bx)) \right )  < \sigma \right )\nonumber \\
& =  P \left (  \sum_{i=1}^\test {W_i}  < \sigma \right )\nonumber\\
& \leq \exp \left (-\frac{\sigma^2/2}{\sum_i^\test EW_i^2 - w\sigma/3} \right ) \nonumber \\
& < \exp \left (-\frac{(\beta_{LP}\log(\bl)(1-2\defp)/e)^2/2}{w^2\beta_{LP}\log(\bl) + w\beta_{LP}\log(\bl)(1-2\defp)/3} \right ) \nonumber \\
& = \exp \left (-\frac{\beta_{LP}\log(\bl)(1-2\defp)^2}{2e^2w(w + (1-2\defp)/3)} \right ), \label{eq:perrbnd}
\end{align}
with $w =  1  + \frac{(1-2\defp)}{\dbnd}$.

We now note that (\ref{eq:perrbnd}) gives an upper bound on the probability that a single perturbation vector in $\Prt'$ causes a non-positive perturbation in optimal value of the objective function of \nlpm.  But there are $\dfct(\bl-\dfct)$ vectors in $\Prt'$. We take a union bound over all of these vectors by multiplying the terms in (\ref{eq:perrbnd}) by $\dfct(\bl-\dfct)$. Hence the overall bound on the probability that any vector from the set $\Prt'$ causes a non-positive perturbation in optimal value of the objective function of \nlpm is given by
\begin{align}
\dfct(\bl-\dfct)\left (
\bl^{ -\frac{\beta_{LP}(1-2\defp)^2}{\ln(2) 2e^2w(w + (1-2\defp)/3)} } \right ) \!<\! \bl^{1 + \Gamma -\frac{\beta_{LP}(1-2\defp)^2}{\ln(2) 2e^2w(w + (1-2\defp)/3)} },
\label{eq:prberr}
\end{align}
with $w =  1  + \frac{(1-2\defp)}{\dbnd}$, where we recall $\Gamma$ is defined as $\ln(\dbnd)/\ln(\bl)$.

The quantity in (\ref{eq:prberr}) bounds from above the probability that \nlpm ``behaves badly" 
({\it i.e.}, some vector from $\Prt'$ causes a non-positive perturbation in optimal value of the objective function of \nlpm).

%
%

We then choose $\beta_{LP}$ as 
\begin{align}
\frac{(\logerror + 1 + \Gamma)\ln(2) 2e^2w(w + (1-2\defp)/3)}{(1-2\defp)^2},
\label{eq:betaexp}
\end{align}
with $w =   1  + \frac{(1-2\defp)}{\dbnd}$, as in the statement of Theorem~\ref{thm:NCBP-LP}. This choice guarantees that the RHS of (\ref{eq:prberr}) is less than $\bl^{-\logerror}$. This shows that with probability at least $1-\bl^{-\logerror}$ {\it all} vectors in $\Prt'$ cause a {\it strictly positive} change in the optimal value of the objective function of \nlpm. 

Finally, we note that the set of feasible $(\bar{\bx},\cost)$ of  \nlpm forms a convex set. Hence if $\cost$ strictly increases along every direction in $\Prt'$, then by Claim~\ref{clm:xset} in fact $\cost$ strictly increases when the true $\bx$ is perturbed in {\it any} direction (in particular, when $\bx$ is perturbed by a strictly positive linear combination of the vectors in $\Prt'$). Hence the true $\bx$ must be the solution to \nlpm.
\hfill $\blacksquare$

\vspace{0.1in}

{
\noindent {\bf Proof of Theorem~\ref{thm:asym}:}\\

Recall that in the asymmetric noise model the probability of false positive and negative test outcomes is denoted by $\defpn$ and $\defpp$ respectively. When $\defpn$ and $\defpp$ are not equal, Equations (\ref{eq:costch2a}) -- (\ref{eq:betaexp}) change as follows. Equation (\ref{eq:costch2a}) becomes 
\begin{align}
& P(\Delta'_{0,i}=1)  =  \pmat(1-\pmat)\left [ (1-\defpn-\defpp)(1-\pmat)^{\dfct-1}+\defpp \right ], \\
& P(\Delta'_{0,i}=-1) =  \pmat(1-\pmat)\defpp.\nonumber
\end{align}
Hence the expected value of the objective value perturbation $\Delta^\test$ equals $\test\pmat(1-\pmat)^{\dfct}(1-\defpn-\defpp)$, and the expected value of the number of perturbed variables $(\#\Delta)^\test$ equals $\test\pmat \left [(1-\pmat)^{\dfct}(1-\defpn-\defpp)+2(1-\pmat)\defpp  \right ]$. These differences lead to changes in the parameters we use in Bernstein's inequality. Specifically, (\ref{eq:wiub}) is now bounded as
\begin{align}
|W_i| = w & =  |\cost_i(\bx')-\cost_i(\bx) - E(\cost_i(\bx')-\cost_i(\bx))|\nonumber\\
& \leq  |\cost_i(\bx')-\cost_i(\bx)|  +  |E(\cost_i(\bx')-\cost_i(\bx))|\nonumber \\
& =  1  + \pmat(1-\pmat)^\dfct(1-\defpn-\defpp)\nonumber \\
& <  1  + (1-\defpn-\defpp)/\dbnd \label{eq:wiub_asym}.
\end{align} 
\noindent Next, we bound $\sum_{i=1}^{\test}EW_i^2$ from above by
\begin{align}
\sum_{i=1}^{\test}EW_i^2 & \leq  w^2(\#\Delta)^\test \nonumber\\
& =  w^2\test\pmat \left [(1-\pmat)^{\dfct}(1-\defpn-\defpp)+2(1-\pmat)\defpp  \right ] \nonumber\\
& <  w^2\beta_{ASY-LP}\log(\bl) \label{eq:sumewi22_asym}.
\end{align}
\noindent We then set 
\begin{align}
\sigma &= -\test E(\cost_i(\bx')-\cost_i(\bx)) \nonumber \\
       &= -\beta_{ASY-LP}\log(\bl)(1-\pmat)^\dfct(1-\defpn-\defpp), \label{eq:sigma_asym}
\end{align}
\noindent which is bounded by
\begin{align}
\log(\bl)(1-\defpn-\defpp)/e < \frac{|\sigma|}{\beta_{ASY-LP}} <\log(\bl)(1-\defpn-\defpp).
\label{eq:sigmabnd_asym}
\end{align}

Substituting the value of $\sigma$ from (\ref{eq:sigma_asym}), $W_i$ from (\ref{eq:wiub_asym}), noting that $\sigma$ is negative, using Claim~\ref{clm:sanov},
and using bounds for $w$, $\sum_{i=1}^{\test}EW_i^2$ and $\sigma$ respectively from (\ref{eq:wiub_asym}), (\ref{eq:sumewi22_asym}) and (\ref{eq:sigmabnd_asym})
gives us that
\begin{align}
P & \left (  \sum_{i=1}^\test \left ( \cost_i(\bx')-\cost_i(\bx) \right )  < 0 \right ) \nonumber \\
  & =  P \left (  \sum_{i=1}^\test \left ( (\cost_i(\bx')-\cost_i(\bx)) - E(\cost_i(\bx')-\cost_i(\bx)) \right )  < \sigma \right )\nonumber \\
& =  P \left (  \sum_{i=1}^\test {W_i}  < \sigma \right )\nonumber\\
& \leq \exp \left (-\frac{\sigma^2/2}{\sum_i^\test EW_i^2 - w\sigma/3} \right ) \nonumber \\
& < \exp \left (-\frac{\beta_{ASY-LP}(\log(\bl)(1-\defpn-\defpp)/e)^2/2}{w^2\log(\bl) + w\log(\bl)(1-\defpn-\defpp)/3} \right ) \nonumber \\
& = \exp \left (-\frac{\beta_{ASY-LP}\log(\bl)(1-\defpn-\defpp)^2}{2e^2w(w + (1-\defpn-\defpp)/3)} \right ), \nonumber
\end{align}
with $w =  1  + \frac{(1-\defpn-\defpp)}{\dbnd}$.

Hence the overall bound on the probability that any vector from the set $\Prt'$ causes a non-positive perturbation in optimal value of the objective function of  \nlpm is given by
\begin{align}
\dfct(\bl-\dfct)\!\left (
\bl^{ -\frac{\beta_{ASY-LP}(1-\defpn-\defpp)^2}{\ln(2) 2e^2w(w + (1-\defpn-\defpp)/3)} } \right ) \!<\! \bl^{1 + \Gamma -\frac{\beta_{ASY-LP}(1-\defpn-\defpp)^2}{\ln(2) 2e^2w(w + (1-\defpn-\defpp)/3)} },
\label{eq:prberr_asy}
\end{align}
with $w =   1  + \frac{(1-\defpn-\defpp)}{\dbnd}$, where we recall $\Gamma$ is defined as $\ln(\dbnd)/\ln(\bl)$.

The quantity in (\ref{eq:prberr_asy}) bounds from above the probability that \nlpm ``behaves badly" 
({\it i.e.}, some vector from $\Prt'$ causes a non-positive perturbation in optimal value of the objective function of \nlpm).

We can thus bound $\beta_{ASY-LP}$ as 
\begin{align}
\frac{(\logerror + 1 + \Gamma)\ln(2) 2e^2w(w + (1-\defpn-\defpp)/3)}{(1-\defpn-\defpp)^2} \label{eq:betaexp_asym}
\end{align}
with $w =   1  + \frac{2(1-\defpn-\defpp)}{\dbnd}$.
\hfill $\blacksquare$

\vspace{0.1in}

\noindent {\bf Proof of Theorem~\ref{thm:act}:}\\

Recall that in the activation noise model, we denote by $\dil$ the probability of activation noise, {\it i.e.}, the probability that a (single) defective item shows up as non-defective in a test. Correspondingly, we denote by $\dil^v$ the probability of observing a false negative test outcome, given that the test involves say $v$ defective items. Also, we use $\defpn$ to denote the probability of a false positive test outcome.

In this noise model, \nlpm decoding still achieves good performance -- ``small'' probability of error   with an order-optimal number of measurements. Again, as in the asymmetric noise model, the proof is essentially the same as the proof for BSC($\defp$) noise. 
Equations (\ref{eq:costch2a}) -- (\ref{eq:betaexp}) change as follows.
Equation (\ref{eq:costch2a}) becomes
\begin{align*}
P(\Delta'_{0,i}=1)  & =  \pmat(1-\pmat)^{\dfct}(1-\defpn) \\
 & +\pmat(1-\pmat)\sum_{i=1}^{\dfct-1}{{\dfct-1 \choose i}(1-\pmat)^{\dfct-1-i}(\pmat\dil)^i} \\
 & =  \pmat(1-\pmat)\left [ (1-\pmat+\pmat\dil)^{\dfct-1}-(1-\pmat)^{\dfct-1}\defpn \right ], \\
P(\Delta'_{0,i}=-1) & =  \pmat\dil(1-\pmat)\sum_{i=0}^{\dfct-1}{{\dfct-1 \choose i}(1-\pmat)^{\dfct-1-i}(\pmat\dil)^i} \\
 & =  \pmat(1-\pmat)(1-\pmat+\pmat\dil)^{\dfct-1}\dil.
\end{align*}
Hence the expected value of the objective value perturbation $\Delta^\test$ equals $$\test\pmat(1-\pmat)\left [(1-\pmat+\pmat\dil)^{\dfct-1}(1-\dil) - (1-\pmat)^{\dfct-1}\defpn  \right ],$$ and the expected value of the number of perturbed variables $(\#\Delta)^\test$ equals $$\test\pmat(1-\pmat)\left [(1-\pmat+\pmat\dil)^{\dfct-1}(1+\dil) - (1-\pmat)^{\dfct-1}\defpn  \right ].$$ 
Substituting into Bernstein's Inequality, we can bound $\beta_{ACT-LP}$ as 
\begin{align}
\frac{(\logerror + 1 + \Gamma)2\ln(2) 2e^2w\left ( 3w(1+\dil-\defpn)-(1-\dil-\defpn) \right )}{3(1-\dil-\defpn)^2}, \label{eq:betaexp_dil}
\end{align}
with $w =   1  + \frac{1-\dil-\defpn}{\dbnd}$.
\hfill $\blacksquare$
}

\noindent {\bf Proof of Theorem~\ref{thm:CBP-LP}:}\\
We substitute $\defp = 0$ in (\ref{eq:betaexp}) to obtain the corresponding $\beta$ as 
\begin{align}
(\logerror + 1 + \Gamma)\ln(2) 2e^2w'(w' + 1/3),  \mbox{ with } w' =   1  + \frac{1}{\dbnd},
\label{eq:betaexp_clean}
\end{align}
\hfill  $\blacksquare$

{\section{Acknowledgments}
The authors wish to acknowledge useful discussions with Pak Hou Che, who contributed to an early version of this work. We would also like to thank the anonymous reviewers, who have suggested new results, pointed out related work, and in general significantly improved the quality of this work.
}

\bibliographystyle{IEEEtran}
\bibliography{IEEEabrv,./group_testing_bib}

\appendix

\section{Proof of the lower bounds in Theorems~\ref{thm:ITLowBnd} and~\ref{thm:ITNoisyLowBnd}}
\label{sec:lwrbnd}
Many ``usual" proofs of lower bounds for group-testing are combinatorial. Information-theoretic proofs are needed to incorporate the allowed probability of error $\error$ into our lower bounds, we provide. Such proofs were provided in, for instance~\cite{Dya:04}. For completeness, we reprise simple versions of these proofs here.

We begin by noting that $\datvectrv \rightarrow \outvectrv \rightarrow \hat{\outvectrv} \rightarrow \hat{\datvectrv}$ ({\it i.e.} the input vector, noiseless result vector, noisy result vector, and the estimate vector) forms a Markov chain. By standard information-theoretic definitions we have
\begin{eqnarray}
	\entropy(\datvectrv) &=& \entropy(\datvectrv | \hat{\datvectrv}) + \mutualinfo(\datvectrv ; \hat{\datvectrv}) \nonumber
\end{eqnarray}
Since $\datvectrv$ is uniformly distributed over all length-$\universe$ and $\dbnd$-sparse data vectors (since $\dfct$ could be as large as $\dbnd$), $\entropy (\datvectrv) = \log |\mathcal{X}| = \log {\universe \choose \dbnd}$. By Fano's inequality, $\entropy(\datvectrv | \hat{\datvectrv}) \leq 1 + \epsilon \log {\universe \choose \dbnd}$. Also, we have $\mutualinfo(\datvectrv ; \hat{\datvectrv}) \leq \mutualinfo(\outvectrv ; \hat{\outvectrv})$ by the data-processing inequality.
Finally, note that $$I({\hat{\mathbf{Y}}};{{\mathbf{Y}}})  \leq\sum_{i=1}^T \left [ H({\hat{Y}_i}) - H({\hat{Y}_i}|Y_i) \right ]$$
since the first term is maximized when each of the ${\hat{Y}_i}$ are independent, and because the measurement noise is memoryless. For the BSC($\defp$) noise we consider in this work, this summation is at most $T(1-H(\defp))$ by standard arguments.\footnote{This technique also holds for more general types of discrete memoryless noise, such as, for instance, the asymmetric noise model considered in Theorem~\ref{thm:asym} -- for ease of presentation, in this work we focus on the simple case of the Binary Symmetric Channel.}

Combining the above inequalities, we obtain
\begin{eqnarray}
	(1 - \epsilon) \log {\universe \choose \dbnd} & \leq & 1 + \test (1 - \entropy{(q)}) \nonumber
\end{eqnarray}
Also, by standard arguments via Stirling's approximation~\cite{CoverandThomas}, $\log{{\bl}\choose {\dbnd}}$ is at least $\dbnd\log(\bl/\dbnd)$. Substituting this gives us the desired result
\begin{eqnarray}
	\test	& \geq & \frac{1 - \epsilon}{1 - \entropy{(\defp)}} \log {\universe \choose \dbnd} \nonumber \\
			& \geq & \frac{1 - \epsilon}{1 - \entropy{(\defp)}} \dbnd \log \left( \frac{\universe}{\dbnd} \right). \nonumber
\end{eqnarray}
\hfill $\blacksquare$

\begin{IEEEbiography}[{\includegraphics[width=1in,height=1.25in,clip,keepaspectratio]{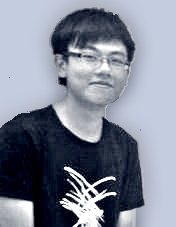}}]{Chun~Lam~Chan}
 received a B.Eng. (2013) in Information Engineering from the Chinese University of Hong Kong (CUHK). He is now a M.Phil. student in the Dept. of Information Engineering in CUHK. His research interest is displayed by the name of the team he is in -- CAN-DO-IT team (Codes, Algorithms, Networks: Design and Optimization for Information Theory).
\end{IEEEbiography}
\begin{IEEEbiography}[{\includegraphics[width=1in,height=1.25in,clip,keepaspectratio]{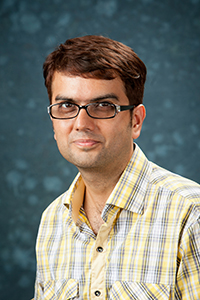}}]{Sidharth Jaggi}
 B.Tech. ('00), EE, IIT Bombay, MS/Ph.D. ('05) EE, CalTech, Postdoctoral Associate ('06) LIDS, MIT, currently Associate Professor, Dept. of Information Engineering, The Chinese University of Hong Kong. More info at http://jaggi.name
\end{IEEEbiography}
\begin{IEEEbiography}[{\includegraphics[width=1in,height=1.25in,clip,keepaspectratio]{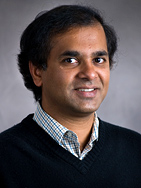}}]{Venkatesh Saligrama}
 (SM'07) is a faculty member in the Electrical and Computer Engineering Department at Boston University. He holds a PhD from MIT. His research interests are in Statistical Signal Processing, Statistical Learning, Video Analysis, Information and Decision theory. He is currently serving as an Associate Editor for IEEE Transactions on Information Theory. He has edited a book on Networked Sensing, Information and Control. He has served as an Associate Editor for IEEE Transactions on Signal Processing and Technical Program Committees of several IEEE conferences. He is the recipient of numerous awards including the Presidential Early Career Award (PECASE), ONR Young Investigator Award, and the NSF Career Award. More information about his work is available at http://blogs.bu.edu/srv
\end{IEEEbiography}
\begin{IEEEbiography}[{\includegraphics[width=1in,height=1.25in,clip,keepaspectratio]{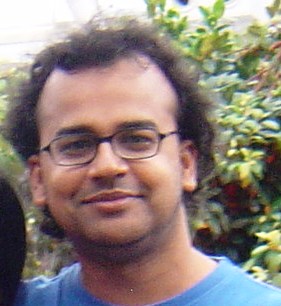}}]{Samar Agnihotri}
 (S'07 - M'10) is a faculty member in the School of Computing and Electrical Engineering at IIT Mandi. He holds M.Sc (Engg.) and Ph.D. degrees in Electrical Sciences from IISc Bangalore. During 2010-12, he was a postdoctoral fellow in the Department of Information Engineering, The Chinese University of Hong Kong. His research interests are in the areas of communication and information theory. More information about his work is available at http://faculty.iitmandi.ac.in/\textasciitilde samar/
\end{IEEEbiography}
\end{document}